\documentclass[lettersize,journal]{IEEEtran}
\usepackage{amsmath,amsfonts}
\usepackage{algorithmic}
\usepackage{algorithm}
\usepackage{array}
\usepackage{color}
\usepackage[caption=false,font=normalsize,labelfont=sf,textfont=sf]{subfig}
\usepackage{geometry}
\usepackage{textcomp}
\usepackage{amssymb}
\usepackage{stfloats}
\usepackage{url}
\usepackage{verbatim}
\usepackage{graphicx}
\usepackage{cite}
\usepackage{bm}
\usepackage{lipsum}
\usepackage{makecell}
\usepackage{longtable}
\usepackage{booktabs}
\usepackage{float}
\usepackage {diagbox}
\usepackage{multirow}
\begin{document}

\title{Knowledge and Data Dual-Driven Channel Estimation and Feedback for Ultra-Massive MIMO Systems under Hybrid Field Beam Squint Effect}

\author{Kuiyu Wang, Zhen Gao, Sheng Chen,~\IEEEmembership{Fellow,~IEEE}, Boyu Ning,~\IEEEmembership{Member,~IEEE}, Gaojie Chen,~\IEEEmembership{Senior Member,~IEEE}, Yu Su,~\IEEEmembership{Senior Member,~IEEE}, Zhaocheng Wang,~\IEEEmembership{Fellow,~IEEE} and H. Vincent Poor, \IEEEmembership{Life Fellow,~IEEE}

\thanks{
	This paper was presented in part at the IEEE 23rd International Conference on Communication Technology (ICCT) [1]. The work of Zhen Gao was supported in part by the Natural Science Foundation of China (NSFC) under Grant 62071044 and Grant U2001210; in part by the Shandong Province Natural Science Foundation under Grant ZR2022YQ62; in part by the Beijing Nova Program. The work of H. Vincent Poor was supported in part by U.S National Science Foundation under Grants CNS-2128448 and ECCS-2335876. \emph{(Corresponding author: Zhen Gao.)}

	Kuiyu Wang and Zhen Gao are with Beijing Institute of Technology (BIT), Beijing 100081, China (e-mails: 3220215125@bit.edu.cn; gaozhen16@bit.edu.cn). 
	
	Sheng Chen is with the School of Electronics and Computer Science, University of Southampton, SO17 1BJ Southampton, U.K. (e-mail: sqc@ecs.soton.ac.uk). 
	
	Boyu Ning is with the National Key Laboratory of Wireless Communications, University of Electronic Science and Technology of China, Chengdu 611731, China (e-mail: boydning@outlook.com).
	
	 Gaojie Chen is with the 5GIC \& 6GIC, Institute for Communication Systems, University of Surrey, GU2 7XH Guildford, U.K. (e-mail: gaojie.chen@surrey.ac.uk).
	 
	  Yu Su is with the China Mobile Chengdu Institute of Research and Development, Chengdu 610000, China (e-mail: suyu@cmii.chinamobile.com).
	  
	   Zhaocheng Wang is with the Department of Electronic Engineering, Tsinghua University, Beijing 100084, China, and also with the Shenzhen International Graduate School, Tsinghua University, Shenzhen 518055, China (e-mail: zcwang@tsinghua.edu.cn).
	   
 H. Vincent Poor is with the Department of Electrical and Computer Engineering, Princeton University, Princeton, NJ 08544 USA (e-mail: poor@princeton.edu).}
\vspace*{-10mm}} %

\maketitle

\begin{abstract}
Acquiring accurate channel state information (CSI) at an access point (AP) is challenging for wideband millimeter wave (mmWave) ultra-massive multiple-input and multiple-output (UM-MIMO) systems, due to the high-dimensional channel matrices, hybrid near- and far- field channel feature, beam squint effects, and imperfect hardware constraints, such as low-resolution analog-to-digital converters, and in-phase and quadrature imbalance. To overcome these challenges, this paper proposes an efficient downlink channel estimation (CE) and CSI feedback approach based on knowledge and data dual-driven deep learning (DL) networks. Specifically, we first propose a data-driven residual neural network de-quantizer (ResNet-DQ) to pre-process the received pilot signals at user equipment (UEs), where the noise and distortion brought by imperfect hardware can be mitigated. A knowledge-driven generalized multiple measurement vector learned approximate message passing (GMMV-LAMP) network is then developed to jointly estimate the channels by exploiting the approximately same physical angle shared by different subcarriers. In particular, two wideband redundant dictionaries (WRDs) are proposed such that the measurement matrices of the GMMV-LAMP network can accommodate the far-field and near-field beam squint effect, respectively. Finally, we propose an encoder at the UEs and a decoder at the AP by a data-driven CSI residual network (CSI-ResNet) to compress the CSI matrix into a low-dimensional quantized bit vector for feedback, thereby reducing the feedback overhead substantially. Simulation results show that the proposed knowledge and data dual-driven approach outperforms conventional downlink CE and CSI feedback methods, especially in the case of low signal-to-noise ratios.
\end{abstract}

\vspace*{-3mm}
\section{Introduction}\label{S1}

Massive multiple-input multiple-output (MIMO) is a key technique to significantly improve spectral efficiency and energy efficiency, owing to its good angular-domain resolution and large array gains in the fifth generation (5G) cellular networks \cite{9711564,8241348}. To provide seamless human-to-everything interactions, the sixth generation (6G) communication system is expected to achieve even higher spectral efficiency and energy efficiency for supporting reliable ultra-high-definition video delivery, extremely low access latency, and real-time interaction with user equipment (UEs) \cite{9798771}. As one of the most disruptive evolution techniques in 6G, ultra-massive MIMO (UM-MIMO) is capable of providing higher flexibility in degrees of freedom and communication capacity. Moreover, by utilizing higher frequency bands, such as millimeter-wave (mmWave) and Terahertz (THz), UM-MIMO is envisioned to support larger transmission bandwidth and shorter latency{ \cite{9237460,BJORNSON20193,add1}}. However, the massive number of antennas results in extremely large array sizes, and the far-field electromagnetic (EM) wave propagation assumption becomes inaccurate \cite{9799524,8844787,9262080}. Moreover, the ultra-wide bandwidth brings the beam squint effects, which is non-negligible in UM-MIMO systems \cite{8354789}. Thus, there is a new urgent need for signal processing algorithms that are aware of and can cope with these challenges.

The acquisition of the downlink channel state information (CSI) at an access point (AP) is particularly challenging for UM-MIMO systems, since estimating the complete CSI at the AP associated with a massive number of antennas at AP would lead to excessive pilot overhead. In the literature, there have been various downlink CSI acquisition schemes proposed for massive MIMO systems \cite{7174558,7355354,gao2014structured,9685542,9598863,8961111,RaoFDDCE}. In time division duplexing (TDD) massive MIMO systems adopting the fully-digital array and serving dozens of UEs, the AP can easily estimate the uplink CSI at an affordable pilot overhead thanks to the relatively limited number of antennas at the UEs as well as the good channel reciprocity between uplink and downlink channels. For conventional sub-6GHz frequency division duplexing (FDD) massive MIMO systems adopting fully-digital arrays, since the uplink and downlink channel reciprocity does not exist, the UEs have to first estimate the downlink CSI based on pilot signals transmitted by the AP and then feed them back to the AP \cite{RaoFDDCE}. However, since the radio frequency calibration in mmWave/THz systems becomes difficult, the uplink and downlink reciprocity in TDD mmWave/THz-based systems deteriorates \cite{7867037}. Moreover, mmWave/THz MIMO systems usually adopt hybrid analog-digital arrays, and the pilot overhead for uplink channels is also proportional to the number of receiver antennas. By contrast, the downlink channel estimation (CE) training time can be relatively smaller, since multiple UEs can simultaneously perform CE according to the downlink pilot signals broadcasted by the AP{\footnote{Since the AP usually has sufficient transmit power, the downlink CE signal-to-noise ratio (SNR) is sufficient to ensure the good estimation accuracy.}}. Furthermore, the sparsity of mmWave/THz MIMO channels can be utilized to substantially reduce CSI feedback overhead. Since for TDD mmWave/THz MIMO systems, it becomes difficult to directly acquire the downlink channels by using the estimated uplink channels with affordable pilot overhead, APs have to ask UEs to perform downlink CE and CSI feedback \cite{9452036}.
\vspace*{-2.8mm}
\subsection{Prior Work}\label{S1.1}

There exists extensive work on the problem of acquiring downlink CSI with affordable pilot overhead. By utilizing the sparsity of mmWave/THz massive MIMO channels represented in the delay domain and/or angle domain, various compressive sensing (CS) based CE algorithms, including greedy and Bayesian inference algorithms, were proposed \cite{7174558,7355354,gao2014structured,9685542,9598863,8961111,8171203}. As typical greedy algorithms, orthogonal matching pursuit (OMP)-type algorithms construct the `best matching' projection of the signal from the redundant measurement matrix in a greedy manner \cite{7174558,7355354,gao2014structured,9685542,9598863}. For example, considering the spatio-temporal common sparsity of delay-domain in FDD massive MIMO systems, the authors in \cite{7355354} proposed a structured CS-based CE scheme, where an adaptive structured subspace pursuit algorithm was developed to improve the CE accuracy. However, the near-field propagation was not considered. To this end, in \cite{9685542}, a simultaneous OMP algorithm was utilized to solve a purely near-field CE problem, where a dedicated dictionary was designed. The CE scheme of \cite{9685542} was further extended to cater to the hybrid near- and far- field channels of UM-MIMO systems in \cite{9598863}, where a variant of OMP was proposed but the wideband communication was not involved. On the other hand, Bayes-based algorithms can utilize \textit{a priori} distribution on sparse matrices to improve CE performance. Specifically, the generalized multiple measurement vector (GMMV) approximate message passing (AMP) algorithm was proposed to realize both active user detection and CE in \cite{8961111}. Considering the practical hardware constraints at the receiver, the authors in \cite{8171203} proposed two CE schemes with low-resolution analog-to-digital converters (ADCs) based on the generalized AMP (GAMP) and vector AMP, respectively. Bayes-based algorithms are able to approach Bayesian optimal performance, but the iterative convergence time is unaffordable in practical implementation.

{In addition to CS-based CE schemes, subspace-based CE schemes exploiting multiple signal classification (MUSIC) or estimation of signal parameters via rotational invariant techniques (ESPRIT) algorithms can directly estimate the dominant channel parameters such as angles of departure/arrival (AoDs/AoAs) and path delays \cite{9684752,9325920,8846224}. In \cite{9325920}, the authors proposed a beam training strategy to measure the AoD and AoA by cooperatively sweeping both wide beams and narrow beams. To deal with the hardware imperfections of low-cost devices, the work \cite{9684752} used the root-MUSIC algorithm in AoD estimation for a receive array with low-resolution ADCs and derived the corresponding Cram\'er-Rao lower bound. The aforementioned research focuses only on incident angle estimation, which neglects parameters such as delay. To this end, in \cite{8846224}, a multi-dimensional unitary ESPRIT algorithm was proposed to estimate AoAs, AoDs, and the corresponding delays at different stages.}

\subsection{Motivation}\label{S1.2}

Although wideband UM-MIMO systems can significantly improve the system throughput to support future 6G networks, several practical challenges beyond the capability of conventional CE schemes need to be addressed.
\begin{enumerate}
	\item {\bf Hardware imperfections}: To support ubiquitous service in future 6G networks, low-cost and energy-saving designs will be widely considered for APs and UEs. Thus, flash ADCs with high sampling speed but moderate resolution are expected to be widely employed in wideband UM-MIMO systems \cite{6189758}. However, this will lead to the received signal suffering from non-negligible information loss. Moreover, in-phase and quadrature (IQ) imbalance will further deepen the distortion of the received signals.
	\item {\bf Hybrid near- and far- field effect}: In conventional MIMO systems, the planar-wave propagation approximation is commonly assumed and the CSI exhibits sparsity in the virtual angular domain. However, the hybrid near- and far- field scenario where far-field and near-field scatterers co-exist is commonly encountered in UM-MIMO systems. Consequently, conventional CE algorithms based on the virtual angular sparsity will work poorly.
	\item {\bf Beam squint effects}: In conventional wideband MIMO systems, the incident virtual angles of EM waves are approximated to be the same across all subcarriers. However, when the system bandwidth increases dramatically, virtual angle offsets appear due to the significant difference in the wavelengths across different subcarriers. If the frequency-flat dictionaries of conventional CE algorithms are still adopted, the virtual angle offsets across different subcarriers will result in severe CE accuracy loss. 
\end{enumerate}
{\renewcommand\arraystretch{1.2}
	\begin{table*}[!tp]
		
		\vspace*{-1mm}
		\caption{Comparison of Existing Deep Learning-based CE Schemes and Proposed Work}
		\label{Tab1}
		\vspace*{-3mm}
		\begin{center}
			\setlength{\tabcolsep}{1mm}{\begin{tabular}{|c|cc|ccc|cc|cc|ccc|}
					\hline
					\multirow{2}{*}{\diagbox{\textbf{\footnotesize\begin{tabular}[c]{@{}c@{}}Research\\ Literature\end{tabular}}}{\textbf{Contents}}} & \multicolumn{2}{c|}{\bf \begin{tabular}[c]{@{}c@{}}\bf Estimation\\ \bf Method\end{tabular}} & \multicolumn{3}{c|}{\bf Drive Type} & \multicolumn{2}{c|}{\begin{tabular}[c]{@{}c@{}}\bf Transmission\\ \bf Direction\end{tabular}} & \multicolumn{2}{c|}{\begin{tabular}[c]{@{}c@{}}\bf Hardware\\ \bf  Imperfection\end{tabular}} & \multicolumn{3}{c|}{\begin{tabular}[c]{@{}c@{}}\bf Hybrid-Field\\ \bf  Beam Squint Effect\end{tabular}} \\ \cline{2-13} 
					
					& \multicolumn{1}{c|}{Subspace} & \begin{tabular}[c]{@{}c@{}}Compressed\\ sensing\end{tabular} & \multicolumn{1}{c|}{\begin{tabular}[c]{@{}c@{}}Data\\ driven\end{tabular}} & \multicolumn{1}{c|}{\begin{tabular}[c]{@{}c@{}}Knowledge\\ driven\end{tabular}}& \multicolumn{1}{c|}{\begin{tabular}[c]{@{}c@{}}Dual\\ driven\end{tabular}} & \multicolumn{1}{c|}{Uplink} & Downlink & \multicolumn{1}{c|}{\begin{tabular}[c]{@{}c@{}}Low-bit\\ ADC\end{tabular}} & \begin{tabular}[c]{@{}c@{}}Other\\ Imperfections\end{tabular} & \multicolumn{1}{c|}{\begin{tabular}[c]{@{}c@{}}Far\\ Field\end{tabular}} & \multicolumn{1}{c|}{\begin{tabular}[c]{@{}c@{}}Near\\ Field\end{tabular}} & \begin{tabular}[c]{@{}c@{}}Hybrid\\ Field\end{tabular} \\ \hline
					
					\cite{9452036} & \multicolumn{1}{c|}{} & \checkmark & \multicolumn{1}{c|}{} & \multicolumn{1}{c|}{\checkmark} & \multicolumn{1}{c|}{}& \multicolumn{1}{c|}{\checkmark} & \checkmark & \multicolumn{1}{c|}{\checkmark} & & \multicolumn{1}{c|}{} & \multicolumn{1}{c|}{}  & \\ \hline    
					
					\cite{9037126} & \multicolumn{1}{c|}{} & \checkmark & \multicolumn{1}{c|}{\checkmark} &\multicolumn{1}{c|}{} & \multicolumn{1}{c|}{}& \multicolumn{1}{c|}{} & \checkmark & \multicolumn{1}{c|}{} & & \multicolumn{1}{c|}{} & \multicolumn{1}{c|}{} & \\ \hline 
					
					\cite{9296779} & \multicolumn{1}{c|}{} & \multicolumn{1}{c|}{\checkmark} & \multicolumn{1}{c|}{}& \multicolumn{1}{c|}{\checkmark} &\multicolumn{1}{c|}{} & \multicolumn{1}{c|}{\checkmark} &  & \multicolumn{1}{c|}{} & \multicolumn{1}{c|}{\checkmark} & \multicolumn{1}{c|}{} & \multicolumn{1}{c|}{} & \\ \hline 
					
					\cite{9246559} & \multicolumn{1}{c|}{} & \multicolumn{1}{c|}{\checkmark} & \multicolumn{1}{c|}{\checkmark} &\multicolumn{1}{c|}{}&\multicolumn{1}{c|}{} & \multicolumn{1}{c|}{\checkmark} & & \multicolumn{1}{c|}{\checkmark} & & \multicolumn{1}{c|}{} & \multicolumn{1}{c|}{} & \\ \hline 	
					
					\cite{9173575} & \multicolumn{1}{c|}{\checkmark} & \multicolumn{1}{c|}{}& \multicolumn{1}{c|}{} & \multicolumn{1}{c|}{\checkmark} &\multicolumn{1}{c|}{}& \multicolumn{1}{c|}{\checkmark} & & \multicolumn{1}{c|}{} & & \multicolumn{1}{c|}{} & \multicolumn{1}{c|}{} & \\ \hline  
					
					\cite{8715473} & \multicolumn{1}{c|}{} & \multicolumn{1}{c|}{\checkmark} & \multicolumn{1}{c|}{} & \multicolumn{1}{c|}{\checkmark}& \multicolumn{1}{c|}{} & \multicolumn{1}{c|}{\checkmark} & & \multicolumn{1}{c|}{} & & \multicolumn{1}{c|}{} & \multicolumn{1}{c|}{} & \\ \hline   
					\cite{9847603} & \multicolumn{1}{c|}{} & \multicolumn{1}{c|}{\checkmark} & \multicolumn{1}{c|}{} & \multicolumn{1}{c|}{\checkmark}& \multicolumn{1}{c|}{} & \multicolumn{1}{c|}{\checkmark} & & \multicolumn{1}{c|}{} & & \multicolumn{1}{c|}{} & \multicolumn{1}{c|}{} & \\ \hline   
					\cite{9685707} & \multicolumn{1}{c|}{} & \multicolumn{1}{c|}{\checkmark} & \multicolumn{1}{c|}{} &  \multicolumn{1}{c|}{\checkmark}& \multicolumn{1}{c|}{} & \multicolumn{1}{c|}{\checkmark} & & \multicolumn{1}{c|}{\checkmark} & & \multicolumn{1}{c|}{\checkmark} & \multicolumn{1}{c|}{} & \\ \hline 	
					
					\cite{DualA1} & \multicolumn{1}{c|}{} & \multicolumn{1}{c|}{\checkmark}  & \multicolumn{1}{c|}{} & \multicolumn{1}{c|}{} & \multicolumn{1}{c|}{\checkmark} & \multicolumn{1}{c|}{\checkmark} & \multicolumn{1}{c|}{} & \multicolumn{1}{c|}{} & \multicolumn{1}{c|}{}& \multicolumn{1}{c|}{} & \multicolumn{1}{c|}{} & \\ \hline
					
					\cite{DualA2} & \multicolumn{1}{c|}{} & \multicolumn{1}{c|}{\checkmark} & \multicolumn{1}{c|}{} &  \multicolumn{1}{c|}{} & \multicolumn{1}{c|}{\checkmark}& \multicolumn{1}{c|}{\checkmark} &\multicolumn{1}{c|}{} & \multicolumn{1}{c|}{} &\multicolumn{1}{c|}{} & \multicolumn{1}{c|}{} & \multicolumn{1}{c|}{} & \\ \hline

					\cite{add_CERIS} & \multicolumn{1}{c|}{} & \multicolumn{1}{c|}{\checkmark}  & \multicolumn{1}{c|}{} & \multicolumn{1}{c|}{}& \multicolumn{1}{c|}{\checkmark}  & \multicolumn{1}{c|}{\checkmark} & \multicolumn{1}{c|}{}& \multicolumn{1}{c|}{} &\multicolumn{1}{c|}{\checkmark} & \multicolumn{1}{c|}{} & \multicolumn{1}{c|}{} & \\ \hline  
					
					Proposed work & \multicolumn{1}{c|}{} & \multicolumn{1}{c|}{\checkmark}& \multicolumn{1}{c|}{} & \multicolumn{1}{c|}{} & \multicolumn{1}{c|}{\checkmark} & \multicolumn{1}{c|}{} & \multicolumn{1}{c|}{\checkmark}  & \multicolumn{1}{c|}{\checkmark} & \multicolumn{1}{c|}{\checkmark} & \multicolumn{1}{c|}{\checkmark} & \multicolumn{1}{c|}{\checkmark} & \multicolumn{1}{c|}{\checkmark} \\ \hline
					
			\end{tabular}}
		\end{center}
		\vspace*{-4mm}
\end{table*}}

With dedicated mathematical modeling and accurate prior knowledge, conventional algorithms have so far performed well in various CE scenarios. However, the performance of conventional algorithms will degrade considerably in envisioned 6G applications due to the environment mismatch and unaffordable time overhead. Recently, deep learning (DL) has been considered as a key enabling technology in various applications of communication systems \cite{VP_book}, e.g., CE \cite{9037126,9296779,9246559,9173575,9452036}, CSI feedback \cite{8322184,8482358,9481880,9662381,9446900}, precoding \cite{9814463}, beam training and prediction \cite{DLbeamTraining1,DLbeamTraining2,DLbeamPrediction}, scheduling \cite{9252919}, and detection \cite{DLdetection1,DLdetection2}. By training a neural network model with a predefined objective function using DL, it can learn features adapted to real-world data, and the trained model can perform prediction in real-time with low complexity. DL-based CE networks can be roughly categorized into data-driven and knowledge-driven according to the adopted mechanisms. In \cite{9037126}, a data-driven end-to-end deep neural network (DNN) was proposed based on the angular sparsity in massive MIMO channels. Further, DL networks have also shown promise in addressing practical issues such as hardware imperfections. For instance, in \cite{9296779}, the authors designed a data-driven CE network to address pilot contamination, synchronization errors, and channel aging. In \cite{9246559}, a conditional generative adversarial network (cGAN) was designed to overcome the challenging quantization noise in a sparse CE problem. 

Beyond these results, and in order to circumvent the overwhelming complexity in conventional iterative algorithms such as sparse Bayesian learning (SBL) and AMP, extensive DL algorithms have been proposed by replacing hyper-parameters with learnable variables, which are known as knowledge-driven algorithms{ \cite{hht_magazine}}. In \cite{8715473}, LampResNet was proposed as a combination of a learned AMP (LAMP) network and a residual neural network (ResNet) to provide coarse and refined estimation results, respectively. Furthermore, the authors in \cite{9452036} unfolded a multiple measurement vector AMP (MMV-AMP) algorithm and designed a learnable phase shifter network at the receiver in an orthogonal frequency domain multiplexing (OFDM) system. By designing a specific shrinkage function, the network leverages the exactly common support across different subcarriers effectively. However, the algorithm requires an identical measurement matrix on different subcarriers and works poorly in the case of hardware imperfections. To address these issues, the authors of \cite{9847603} proposed a few-bit massive MIMO channel estimation network to achieve CE and pilot training at the AP and UEs simultaneously, but the network is limited to narrowband systems and cannot be easily applied to wideband systems. To handle the beam squint effects in mmWave wideband systems, the authors in \cite{9685707} proposed a learnable iterative shrinkage thresholding algorithm-based channel estimator by transforming the beam-frequency mmWave channels into sparse representations, but the impact of near-field propagation was not considered.

In addition to performing the downlink CE, UEs are required to feed the received pilot signal or estimated CSI back to the AP. However, if the CSI matrix is quantized directly and fed back to the AP, it will result in non-negligible quantization noise. To this end, the authors of \cite{8322184} utilized the virtual angular sparsity to improve massive MIMO CSI compression. However, this early work did not consider the impact of existing knowledge at the AP. To this end, in \cite{8482358}, the time correlation of time-varying massive MIMO channels was utilized by introducing a long short-term memory structure, and the network was demonstrated to outperform schemes not considering this effect. Besides, in \cite{9481880}, the authors took advantage of the partial channel reciprocity between uplink and downlink channels in FDD massive MIMO systems. However, the input of the aforementioned schemes is assumed to be noiseless, which is impossible to obtain in practical transmission systems. For this reason, in \cite{9171358} a two-module neural network termed an anti-noise CSI compression network was developed to achieve noisy CSI feedback compression, where the first module removed the noise and the next module compressed the CSI and eliminated the residual noise in the compression. However, the training is not end-to-end in the CSI feedback and training takes a long time. Moreover, the beam squint effects make the essential information extraction more complicated and degrades the performance of the aforementioned schemes. 

In machine learning terminology, purely data-driven modeling is known as black-box modeling, while pure knowledge-driven modeling is referred to as white-box modeling. A fundamental principle in data modeling is to incorporate available \textit{a priori} information, i.e., \textit{a priori} knowledge, regarding the underlying data-generating mechanism into the modeling process. Knowledge-and-data dual modeling, also known as grey-box modeling, is capable of incorporating prior knowledge and typically outperforms purely data-driven modeling \cite{greyboxM1,greyboxM2,greyboxM3}. Thus, a knowledge and data dual-driven approach should offer a much better method for the CE and CSI feedback. However, few studies have considered this approach in the present context.  In \cite{DualA1}, the authors proposed a scheme that includes a data-driven noise level detecting network to aid a knowledge-driven LAMP CE network, while the scheme of \cite{DualA2} includes a knowledge-driven Gaussian mixture LAMP CE network and a data-driven residual learning network for CSI-denoising. However, the aforementioned research is limited to the case of narrowband transmission. {The authors in \cite{add_CERIS} proposed a hybrid driven channel estimation scheme for a reconfigurable intelligent surface aided wideband system, but the wideband beam squint effect and near-field effect are not considered.} Furthermore, the scenarios with near-field propagation and hardware imperfections are not addressed in these schemes either.
\vspace*{-3.0mm}
\subsection{Our Contributions}\label{S1.3}
Against the above background, this paper proposes a novel knowledge and data dual-driven downlink CE and uplink CSI feedback approach for UM-MIMO systems, where an AP serves multiple single-antenna UEs using wideband OFDM transmission. We assume that the scatterers between the UE and the AP can be located in either the near or far field with respect to the AP. The AP adopts hybrid beamforming and all UEs use low-bit ADCs as a cost-effective solution for practical deployment. Our main contributions are summarized as follows.
\begin{itemize}
	\item{\textbf{A data-driven de-quantizer (DQ) module is proposed to combat hardware imperfections.} All UEs exploit the oversampled received pilot signal to mitigate the signal distortion caused by the low-bit ADCs. Specifically, a data-driven ResNet DQ (ResNet-DQ) is developed to mitigate signal distortion caused by quantization and additive white Gaussian noise (AWGN). Simulation results demonstrate that ResNet-DQ can eliminate both AWGN and quantization noise effectively, especially in low-SNR regions. {To the best of our knowledge, this is the first attempt to achieve time-domain OFDM signal de-quantization based on deep learning, where the intrinsic correlation of oversampled signals at a practical receiver is exploited.}}
	\item{\textbf{Two wideband redundant dictionaries (WRDs) are proposed to sparsify UM-MIMO channel matrices.} Two customized WRDs are designed for far-field and near-field, respectively. The former is derived based on the discrete Fourier transform (DFT) matrix, while the latter is obtained by data-driven methods. Both dictionaries aim at compensating for the virtual angle-domain support offsets under beam squint effects, so that the almost identical physical AoA/AoDs across different subcarriers can be exploited.}
	\item{\textbf{A knowledge-driven network is utilized to estimate the channels.} By using the designed WRDs, the sparse angle-domain vectors on different subcarriers exhibit exactly the common sparse support. This enables us to formulate the downlink wideband CE as a sparse signal recovery problem and develop a GMMV-LAMP network to estimate the channels with low pilot overhead. As the de-quantization procedure in ResNet-DQ is highly non-linear, the received signals from different subcarriers suffer from different noise levels and we derive a shrinkage function to deal with this. Simulation results verify the excellent CE performance in a wideband mmWave system with the beam squint effects.}
	\item{\textbf{A data-driven module is proposed to efficiently perform bit-vector CSI feedback.} It consists of a CSI-ResNet auto-encoder with encoding and decoding components deployed at the UEs and AP, respectively. With dedicated designed network structures, the encoder at a UE can compress CSI and transform it into a compact bit-vector, then the AP reconstructs the high-dimensional CSI once the feedback vector is received by a decoder with a similar structure. Simulation results verify this module's effectiveness in CSI feedback especially in low-SNR regions, which outperforms the widely considered CSI-Net \cite{8322184}.}
\end{itemize}

\textit{Notation}: We use lower-case bold letters for vectors, e.g., $\mathbf{a}$, and capital bold letters for matrices, e.g., $\mathbf{A}$. The conjugate, transpose and conjugate transpose operators are denoted by $(\cdot )^{*}$, $(\cdot )^{\rm T}$ and $(\cdot )^{\rm H}$, respectively. The $i$th element of $\mathbf{a}$ is given by $[\mathbf{a}]_{i}$ and the $(i,j)$th element of $\mathbf{A}$ is denoted as $[\mathbf{A}]_{i,j}$. For a diagonal matrix $\mathbf{\Sigma}$, $\mathbf{\Sigma}^{\frac{1}{2}}$ denotes the diagonal matrix in which each diagonal element is the square root of the corresponding element in $\mathbf{\Sigma}$. $\Re (\mathbf{a})$ ($\Re (\mathbf{A})$) and $\Im (\mathbf{a})$ ($\Im (\mathbf{A})$) denote the real part and imaginary part of $\mathbf{a}$ ($\mathbf{A}$), respectively. $\left \lceil a \right \rceil $ rounds $a$ to the nearest integer greater than or equal to $a$. $\mathbf{a}\sim \mathcal{CN} \left(\mathbf{\bar{a}},\mathbf{V}\right)$ denotes a random vector $\mathbf{a}$ following the complex Gaussian distribution with mean vector $\mathbf{\bar{a}}$ and covariance matrix $\mathbf{V}$. $\mathcal{U}\left(a,b\right)$ denotes the uniform distribution within the range of $\left[a,\, b\right]$. $\mathbf{0}$ and $\mathbf{I}$ denote the zero vector and identity matrix of appropriate dimension, respectively. $\left \| \mathbf{a} \right \|_{0}$ denotes the $l_{0}$ norm of $\mathbf{a}$. $\left \| \mathbf{A} \right \|_{F}$ denotes the Frobenius norm of $\mathbf{A}$. $\frac{\partial\left[\mathbf{f}(\mathbf{r})\right]_{j}}{\partial x}$ denotes the partial derivative of the $j$th element of a vector-valued function $\mathbf{f}(\mathbf{r})$ with respect to variable $x$. $\mathbb{E}\left[\cdot \right]$ denotes the statistical expectation operator.
\vspace*{-2mm}
\section{Preliminaries}\label{S2}

We first present the system model which consists of the transmission model and entire neural network structure for CE. Then, the mmWave wideband channel model that considers the hybrid near- and far- field effect and beam squint effect is detailed.

\vspace*{-5.0mm}
\subsection{System Model}\label{S2.1}

We consider the downlink CE and CSI feedback problem in a single-cell multi-user mmWave wideband UM-MIMO system. The carrier frequency and the corresponding wavelength are denoted by $f_{c}$ and $\lambda_{c}$, respectively. The AP adopts the hybrid analog-digital MIMO architecture with $N_{\mathrm{RF}}$ radio frequency (RF) chains and an $N_{\mathrm{AP}}$-element uniform linear array (ULA). The antenna spacing is half of the carrier wavelength, i.e., $d = \lambda_{c}/2$. Without loss of generality, each UE is equipped with a single antenna and a $Q$-bit ADC to reduce hardware cost and power consumption. The center of AP is positioned at $(0,0)$, and the coordinates of antennas are $\left(0,-\frac{\lambda_{c}}{4}+(i-\frac{N_{\mathrm{AP}}}{2})\frac{\lambda_{c}}{2}\right),\forall i\in\{1,...,N_{\mathrm{AP}}\}$. The number of scatterers $L$ is assumed to be the same for all UEs. OFDM modulation with $K$ subcarriers is adopted to combat the multipath channels. 

\begin{figure*}[tp!]
\vspace*{-1mm}
\begin{center}
\includegraphics[width=0.99\textwidth]{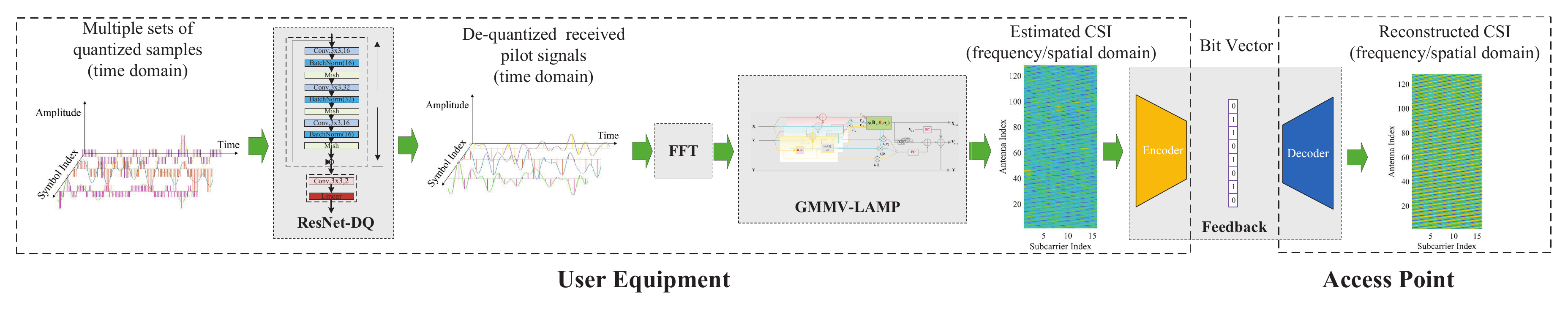}
\end{center}
\vspace*{-4mm}
\caption{Structure of the knowledge and data dual-driven network:  the whole network is divided into data-driven de-quantization, knowledge-driven GMMV-LAMP, and data-driven CSI feedback modules. Time-domain quantized oversampling samples are first inputted into the de-quantization module, and the output is utilized in GMMV-LAMP for CE. Finally, the estimated CSI is compressed and reconstructed in the CSI feedback module.}

\label{system_structure} 
\vspace*{-4mm}
\end{figure*}

We design a transmission frame that includes two stages: the first $G$ pilot slots with $K$ subcarriers are utilized for CE, and the rest $(T-G)$ slots with $S_{d}$ subcarriers are used for data transmission. Note that the pilot symbol length is much smaller than the data symbol length, and the CE overhead is relatively small compared with the data transmission, in which the subcarrier spacing becomes larger. Assume that the maximum number of delay taps is $\left\lceil \tau_{\mathrm{max}} f_{s} \right\rceil$, and the number of pilot subcarriers satisfies $K> \left\lceil \tau_{\mathrm{max}} f_{s} \right\rceil $, where $\tau_{\mathrm{max}}$ denotes the maximum delay of the wideband channels. At the CE stage, the AP broadcasts the pilot signals to all UEs. After the received signal is quantized with a low-bit ADC, it is inputted into the proposed network illustrated in Fig.~\ref{system_structure}. Specifically, it is first de-quantized by ResNet-DQ. Then, the recovered high-resolution signal is utilized for CE using GMMV-LAMP. After the UE obtains the complete CSI estimated by GMMV-LAMP, a dedicated encoder compresses the CSI matrix into a low-dimensional bit vector which is then fed back to the AP{\footnote{As different UEs have the same network structure, the time overhead of CE and CSI compression for all UEs should be similar. Thus, we assume that all UEs feed their own bit vectors back to the AP simultaneously for convenience.}}. Finally, the AP reconstructs the CSI of UEs with an identical decoder. It is worth noting that UEs' propagation environments are similar. Without loss of generality, we can focus on the CE and CSI feedback of a single UE, and indices of UEs can be omitted.

Denote the signal transmitted by the AP on the $k$th subcarrier and the $g$th pilot slot by $\mathbf{F}_{\mathrm{RF}}[g]\mathbf{s}[g,k]\in \mathbb{C}^{N_{\mathrm{AP}}\times 1}$, where $\mathbf{F}_{\mathrm{RF}}[g]=\frac{1}{\sqrt{N_{\mathrm{AP}}}}\bm{e}^{\textsf{j}\bm{\Xi}[g]} \in \mathbb{C}^{N_{\mathrm{AP}}\times N_{\mathrm{RF}}}$ and $\mathbf{s}[g,k]\in \mathbb{C}^{N_{\mathrm{RF}}\times 1}$ denote the analog precoder on the $g$th pilot slot and the baseband pilot symbol on the $k$th subcarrier and the $g$th pilot slot, respectively. The elements $\xi_{i,l}[g]=[\bm{\Xi}[g]]_{i,l}$ for $1\le i\le N_{\mathrm{AP}}$ and $1\le l\le N_{\mathrm{RF}}$ are uniformly distributed in $\mathcal{U}(0, 2\pi)$ and the $(i,l)$th element of the precoding matrix $\bm{e}^{\textsf{j}\bm{\Xi}[g]}$ is $e^{\textsf{j}\xi_{i,l}[g]}$. Furthermore, $\mathbf{s}[g,k]\sim \mathcal{CN}(\mathbf{0},\mathbf{I})$. To avoid the peak-to-average power ratio issue, $\mathbf{s}[g,i]\neq \mathbf{s}[g,j],\forall i,j\in\{1,2,...,K\}$ and $i\neq j$. At the UE, the received signal under the multipath channels on the $k$th subcarrier and the $g$th pilot slot can be formulated as 
\begin{equation}\label{system model 1} 
	\vspace*{-2mm}
  \check{y}_{\mathrm{DL}}[g,k] = \mathbf{h}_{\mathrm{DL}}^{{\rm T}}[k] \mathbf{F}_{\mathrm{RF}}[g] \mathbf{s}[g,k] + \check{n}_{\mathrm{DL}}[g,k] ,
\end{equation}
where $\mathbf{h}_{\mathrm{DL}}[k]\! \in\! \mathbb{C}^{N_{\mathrm{AP}}\times 1}$ denotes the downlink channel on the $k$th subcarrier, and the frequency-domain channel AWGN $\check{n}_{\mathrm{DL}}[g,k]\! \sim\! \mathcal{CN} \left(0,\check{\sigma}_n^2\right)$, whose power $\check{\sigma}_n^2$ is frequency-independent and time invariant. Collecting the signals on all subcarriers into $\check{\mathbf{y}}_{\mathrm{DL}}[g]\! =\! \left[\check{y}_{\mathrm{DL}}[g,1],\ldots,\check{y}_{\mathrm{DL}}[g,K]\right]^{\rm T}\! \in\! \mathbb{C}^{K\times 1}$, the received time-domain samples $\check{\mathbf{y}}^{t}_{\mathrm{DL}}[g]\in \mathbb{C}^{K\times 1}$ can be written as 
\begin{equation}\label{system model 2} 
	\vspace*{-2mm}
  \check{\mathbf{y}}^{t}_{\mathrm{DL}}[g] = \mathbf{F}_{\mathrm{DFT},K}^{\rm H} \check{\mathbf{y}}_{\mathrm{DL}}[g],
\end{equation}
where $\mathbf{F}_{\mathrm{DFT},K}$ represents the $K$-dimensional DFT matrix. 
\begin{figure}[bp!]
	\vspace*{-6mm}
	\begin{center}
		\includegraphics[width=0.48\textwidth]{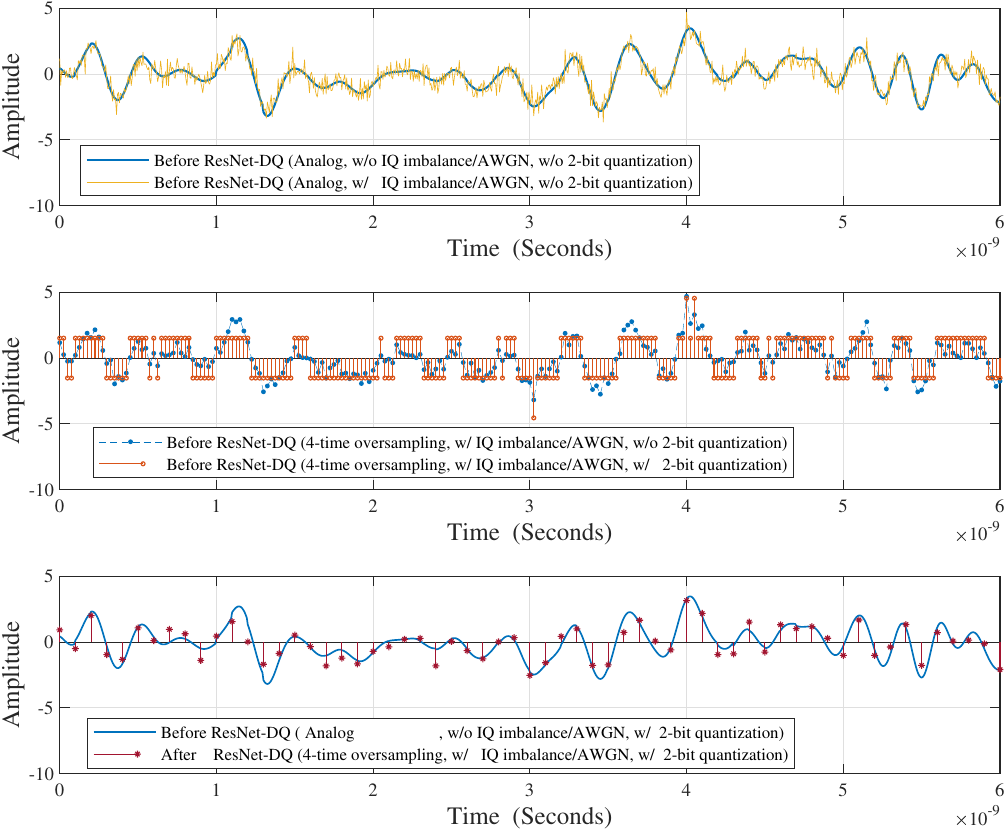}
	\end{center}
	\vspace*{-5mm}
	\caption{Illustration of the de-quantization: the top subfigure depicts the analog perfect received signal without noise or distortion and the analog distorted received signal; the middle subfigure depicts the sampled distorted received signal and its quantized version under 4-time oversampling; the bottom subfigure compares the analog perfect received signal with the sampled signal after ResNet-DQ, which demonstrates that most of the information has been recovered.}
	\vspace*{-6mm}
	\label{Upsampling} 
\end{figure}

However, in practical systems there exist several factors that can cause the distortion of the received signal, e.g.,  IQ imbalance and low-resolution ADC. Denote the $g$th perfect analog baseband signal vector by $\check{\mathbf{y}}_{\mathrm{DL}}(g,t)$, which is generated by $\check{\mathbf{y}}^{t}_{\mathrm{DL}}[g]$ with the aid of shaping filters. As described in \cite{950789}, the analog baseband signal with IQ imbalance $\mathbf{y}_{\mathrm{DL,IQ}}(g,t)$ can be formulated as
\begin{align}\label{IQimbalance} 
	\mathbf{y}_{\mathrm{DL,IQ}}(g,t) &= \left(\cos{\big(\zeta_{\theta}/2\big)} + \textsf{j}\zeta_{A}\sin{\big(\zeta_{\theta}/2\big)}\right) \check{\mathbf{y}}_{\mathrm{DL}}(g,t) \nonumber\\ & + \left(\zeta_{A}\cos{\big(\zeta_{\theta}/2\big)} - \textsf{j}\sin{\big(\zeta_{\theta}/2\big)}\right) \check{\mathbf{y}}_{\mathrm{DL}}^{*}(g,t) ,
\end{align}
where $\zeta_{A}$ and $\zeta_{\theta}$ denote the gain and phase error factors, respectively, $\mathbf{y}_{\mathrm{DL,IQ}}(g,t)$ is sent to an $Q$-bit ADC to obtain the $Q$-bit quantized signal $\mathbf{Y}^{t}_{\mathrm{DL}}[g]\in\mathbb{C}^{W\times K}$ with $W$-times oversampling:
\begin{equation}\label{system model 2.5} 
	\mathbf{Y}^{t}_{\mathrm{DL}}[g] = \mathcal{Q}\left(\mathbf{y}_{\mathrm{DL,IQ}}\left(g,t\right);W,Q\right),
\end{equation} 
where $\mathcal{Q}(\cdot ;W,Q)$ denotes the complex-valued $Q$-bit quantization function. As illustrated in the top and middle subfigures of Fig.~\ref{Upsampling}, the analog baseband signal is quantized by $Q$-bit with $W$ times oversampling. Specifically, each element of $\Re (\mathbf{y}_{\mathrm{DL,IQ}}(g,t))$ and $\Im (\mathbf{y}_{\mathrm{DL,IQ}}(g,t))$ is approximated by the closest value within the quantized set $\mathcal{C}_{b}$, which consists of the $2^{Q}$ candidates, given by
\begin{equation}\label{system model 3} 
	\mathcal{C}_{b}: \left\{-\frac{2^{Q-1}}{2}\Delta_{b},\bigg(-\frac{2^{Q-1}}{2}+1\bigg)\Delta_{b},\ldots ,\frac{2^{Q-1}}{2}\Delta_{b}\right\}  ,
\end{equation}
where $\Delta_{b} = \frac{1}{2^{Q}} \left( Y_{\max} - Y_{\min}\right)$ with $Y_{\max} =\!\!\! \max\limits_{1\le i\le W, 1\le l\le K}\!\! \left\{\Re\big(\big[\mathbf{Y}_{\mathrm{DL},{\rm IQ}}(g,t)\big]_{i,l} \big) ,\Im\big(\big[\mathbf{Y}_{\mathrm{DL},{\rm IQ}}(g,t)\big]_{i,l} \big)\right\}$ and $Y_{\min}\! =\!\!\!\! \min\limits_{1\le i\le W, 1\le l\le K}\!\! \left\{\Re\big(\big[\mathbf{Y}_{\mathrm{DL},{\rm IQ}}(g,t)\big]_{i,l} \big) ,\Im\big(\big[\mathbf{Y}_{\mathrm{DL},{\rm IQ}}(g,t)\big]_{i,l} \big)\right\}$. 
\vspace*{-4.0mm}
\subsection{Channel Model}\label{S2.2}

\begin{figure}[b!]
		\centering
		\includegraphics[scale=0.4]{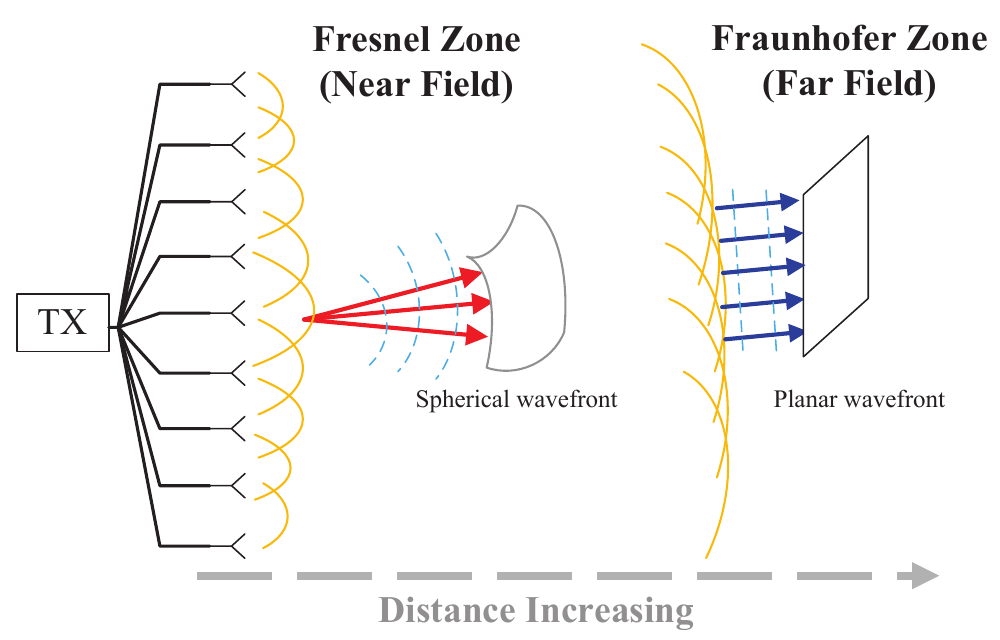}
		\caption{The transmission region can be split into far-field and near-field regions.}
		\label{planar_spherical}

\end{figure}

In UM-MIMO systems, the large array aperture leads to a significant increase in the Rayleigh distance, causing the transmission region to be split into far-field and near-field regions, as illustrated in Fig.~\ref{planar_spherical}. We first present the near-field channel model and then introduce the far-field channel model. Next, the impact of the hybrid near- and far- field scenario is discussed, and the beam squint effects caused by a very large bandwidth are considered.

\subsubsection{Near-Field Channel Model}

\begin{figure}[t!]
\vspace*{-4mm}
	\begin{minipage}[t]{0.45\linewidth}
		\centering
		
		\includegraphics[scale=0.45]{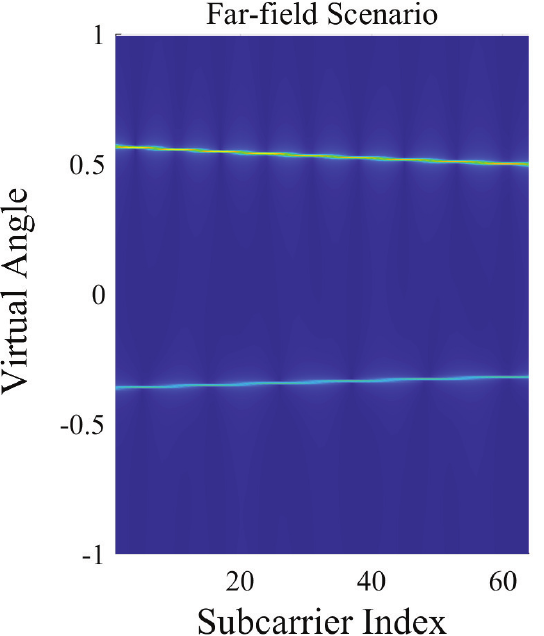}

		\caption{Far-field beam squint effect for the case of 2 multipath components.}
		\label{far_squint}
	\end{minipage}
	\hfill
	\begin{minipage}[t]{0.45\linewidth}
		\centering
		\includegraphics[scale=0.45]{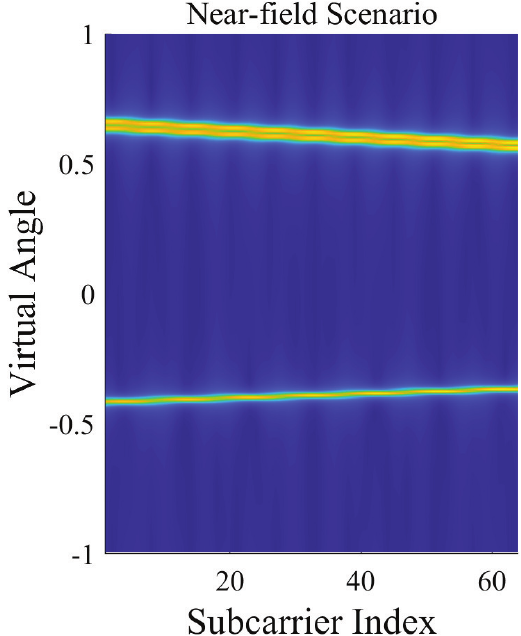}

		\caption{Near-field beam squint effect for the case of 2 multipath components.}
		\label{near_squint}
	\end{minipage}
\vspace*{-4mm}
\end{figure}

For the near-field scenario, the channel model can be derived based on the accurate spherical-wave propagation model. According to the mmWave multipath Saleh-Valenzuela channel model \cite{svchannel}, if there is no angle spread and only one ray for each scatterer, the frequency-domain channel between the $i$th antenna at the AP and the UE on the $k$th subcarrier, $h_{\mathrm{DL},i}[k]=\big[\mathbf{h}_{\mathrm{DL},i}[k]\big]_i$, can be written as
\begin{equation}\label{CM_1 FAR} 
  h_{\mathrm{DL},i}[i] = \sqrt{\frac{1}{LN_{\mathrm{AP}}}} \sum_{l=1}^{L} \beta_{l} e^{\frac{-\textsf{j}2\pi d_{l,i}}{\lambda_{k}}} ,
\end{equation}
where $\beta_{l}\sim \mathcal{CN}(0,1)$ is the channel gain of the $l$th path, $\lambda_{k}$ denotes the wavelength of the $k$th subcarrier, and $d_{l,i}=\sqrt{x_{l}^{2}+\big(-\frac{\lambda_{c}}{4}+\big(i-\frac{N_{\mathrm{AP}}}{2}\big) \frac{\lambda_{c}}{2}-y_{l}\big)^{2}}$ defines the distance of the $l$th path between the UE and the $i$th antenna at the AP with $\left(x_{l},y_{l}\right)$ denoting the Cartesian coordinate of the $l$th scatterer. Collecting the channel response associated with all the antennas of the AP, the channel vector on the $k$th subcarrier, $\mathbf{h}_{\mathrm{DL}}[k] = \left[h_{\mathrm{DL},1}[k],\ldots ,h_{\mathrm{DL},N_{\mathrm{AP}}}[k]\right]^{\rm T}$, can be written as 
\begin{equation}\label{CM_1 NEAR} 
  \mathbf{h}_{\mathrm{DL}}[k] = \sqrt{\frac{1}{LN_{\mathrm{AP}}}} \sum_{l=1}^{L} \beta_{l} e^{\frac{-\textsf{j}2\pi kf_{s}\tau_{l}}{K}}\mathbf{a}_{(x_{l},y_{l})}[k],
\end{equation}
where $\tau_{l}$ represents the delay of the $l$th path associated with the reference antenna, and $f_{s}$ is the system bandwidth, while $\mathbf{a}_{(x_{l},y_{l})}[k]$ is the near-field array steering vector for the $l$th path on the $k$th subcarrier, which can be written as
\begin{equation}\label{near-field vector} 
  \mathbf{a}_{(x_{l},y_{l})}[k] = \Big[e^{-\textsf{j}\frac{2\pi D_{1,l}}{\lambda_{k}}},\ldots,e^{-\textsf{j} \frac{2\pi D_{i,l}}{\lambda_{k}}},\ldots,e^{-\textsf{j}\frac{2\pi D_{N_{\mathrm{AP}},l}}{\lambda_{k}}}\Big]^{\rm T},
\end{equation}
where $D_{i,l}=d_{i,l}-d_{1,l}$ defines the relative distance of the $l$th path between the $i$th antenna and the UE. 
\begin{figure}[bp!]
	\vspace*{-4mm}
	\begin{center}
		\includegraphics[width=0.35\textwidth]{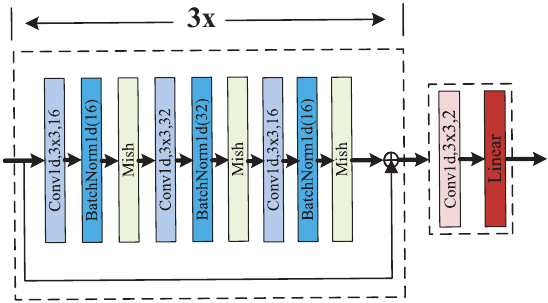}
	\end{center}
	\vspace*{-3mm}
	\caption{Structure of ResNet-DQ.}
	\label{dequan_block} 
	
\end{figure}
\subsubsection{Far-Field Channel Model}

The far-field channel model is based on the planar-wave approximation. When the path distances between the UE and AP are relatively large, the phase differences across different antennas can be determined merely by AoDs. Let the AoD of the $l$th far-field path be $\varphi_{l}=\arctan{(\frac{y_{l}}{x_{l}})}$. Then the array steering vector \eqref{near-field vector} can be approximated by 
\begin{equation}\label{farfield-steeringvector} 
	\mathbf{a}_{\sin{(\varphi_{l})}}[k]\! =\! \Big[1,e^{-\textsf{j}\frac{\pi \lambda_{c} \sin{(\varphi_{l})}}{\lambda_{k}}},\ldots ,e^{-\textsf{j}\frac{(N_{\mathrm{AP}}-1)\pi \lambda_{c} \sin{(\varphi_{l})}}{\lambda_{k}}}\Big]^{\rm T}\!\! .\!
\end{equation}
When the path distances are larger than the Rayleigh distance, the channel vector on the $k$th subcarrier $\mathbf{h}_{\mathrm{DL}}[k]$ can be regarded as a function solely associated with the AoDs of all scatterers. In this case, $\mathbf{h}_{\mathrm{DL}}[k]$ exhibits sparsity in the virtual angular domain, which can be exploited in CE.

\subsubsection{Hybrid Near- and Far- Field Channel Model}

In UM-MIMO systems, the Rayleigh distance can reach dozens or even hundreds of meters, and scatters can be located in both the near-field and far-field regions of the AP. Hence, both far-field scatterers and near-field scatterers coexist, which is termed a hybrid near- and far- field effect in channel modeling \cite{9598863}. {In such a scenario, the channel vector on the $k$th subcarrier can be expressed as \eqref{CM_1 Hybrid} which is located at the top of the next page,} where $L_{n}$ and $L$ denote the number of near-field and all scatterers, respectively. $L_{f}=L-L_{n}$ denotes the number of far-field scatterers. As the near-field scatterers are determined by the Cartesian coordinates of scatterers rather than solely by AoDs, the CE system that is based on the angular sparsity will suffer severe performance loss. Note that the array steering vectors in \eqref{near-field vector} are related to the Cartesian coordinates of scatterers, which can be transformed into the AoD-distance pairs. This transformation plays a crucial role in our data-driven dictionary design discussed in the next section.

\begin{figure*}[t!]
{
\begin{equation}
\label{CM_1 Hybrid} 
	\mathbf{h}_{\mathrm{DL}}[k] = \sqrt{\frac{1}{LN_{\mathrm{AP}}}} \Bigg( \underbrace{\sum_{l=1}^{L_{n}} \beta_{l} e^{\frac{-\textsf{j}2\pi k f_{s}\tau_{l}}{K}} \mathbf{a}_{(x_{l},y_{l})}[k] }_{\mathrm{near-field\  part}}
	+ \underbrace{\sum_{l=L_{n}+1}^{L} \beta_{l} e^{\frac{-\textsf{j}2\pi k f_{s}\tau_{l}}{K}} \mathbf{a}_{\sin{(\varphi_{l})}}[k]}_{\mathrm{far-field\  part}}\Bigg).
\end{equation}}
\rule{\textwidth}{1pt}
\vspace*{-8mm}
\end{figure*}

\subsubsection{Beam Squint Effects}

In narrowband systems, all the subcarriers have almost the same wavelength, and the antenna spacing is designed to be the half wavelength of the center subcarrier. Consequently, for both the near-field and far-field cases, the array steering vector is regarded to be the same for all the subcarriers, that is, $\mathbf{a}_{(x_{l},y_{l})}[k]$ and $\mathbf{a}_{\psi_{m}}[k]$ are identical for all the subcarrier indices $k$. However, for very large wideband systems with a bandwidth comparable to the carrier frequency, the array steering vectors are frequency-selective in both the near-field and far-field cases. This phenomenon is commonly known as the beam squint effects, {which is also called as beam split effect in \cite{cuibeamsplit,wang2019beamsquint,10130575}}. When the frequency-flat dictionary is employed in MMV sparse signal recovery algorithms, the virtual angular-domain support shifts across different subcarriers as illustrated in Fig.~\ref{far_squint} and Fig.~\ref{near_squint}, respectively. Consequently, this phenomenon will lead to severe performance loss.

\section{Knowledge and Data Dual-Driven CE and Feedback Network}\label{S3}

\subsection{Data-Driven De-quantization Module}\label{S3.1}

Recall that in most receivers with high-speed low-resolution ADCs, multiple times of time-domain oversampling and phase-locked loops are applied to identify the optimal sampling moment. { In this case, there is a correlation between different branches of the oversampled signal, which is retained even after quantization. Since it is difficult to design a knowledge-driven network for utilizing the correlation above accurately, we propose a data-driven de-quantization network called ResNet-DQ as depicted in Fig.~\ref{dequan_block}.} ResNet-DQ is composed of three identical ResNet blocks and a CNN dimension adjustment layer, in which we can take advantage of the time-domain correlations across multiple sets of quantized samples. Based on this feature, how to recover the noiseless signal is equivalent to a classic super-resolution problem, in which multiple sets of distorted copies are used to reconstruct the target signal. For convenience, we use quadruple oversampling, i.e., $W=4$, as an illustrative example.

The UE first removes the cyclic prefix of each OFDM symbol. Let the $g$th OFDM symbol be the input to ResNet-DQ. As DL networks do better in real-valued computations, we stack the real and imaginary parts of the quadruple samples of the $g$th OFDM symbol, denoted by $\mathbf{y}_{\mathrm{DL,ov},g}\in \mathbb{C}^{4\times K}$, as an $8$-channel input $\mathbf{y}_{\mathrm{DL,ov,re},g}\in \mathbb{C}^{8\times K}$, and set the output channels of the ResNet block to $8$. For the CNN layer, the input and output channels are $8$ and $2$, respectively, and its $2$-channel output signal $\mathbf{y}_{\mathrm{DL,re},g}\in \mathbb{C}^{2\times K}$ is recast as a complex-valued vector $\mathbf{y}^{t}_{\mathrm{DL},g}\in \mathbb{C}^{K\times 1}$. Finally, $G$ OFDM symbols are collected and reshaped as a matrix $\mathbf{Y}^{t}_{\mathrm{DL},g}\in \mathbb{C}^{G\times K}$, which is then transformed into the frequency domain signal $\mathbf{Y}_{\mathrm{DL}}$ for wideband CE. To make ResNet-DQ fulfill de-quantization well, we design a training strategy which is detailed in Subsection~\ref{S3.4}. As the bottom subfigure of Fig.~\ref{Upsampling} shows, although there still exists some residual noise in the time-domain signal after ResNet-DQ, most of the essential information has been recovered, which significantly contributes to the accuracy of CSI acquisition.

\subsection{Knowledge-Driven CE Module}\label{S3.2}

By considering quantization and de-quantization, the transmission model (\ref{system model 1}) for $1\le g\le G$ can be collected together as
\begin{equation}\label{CE1} 
	\mathbf{y}_{\mathrm{DL}}[k] = \mathbf{S}[k] \mathbf{h}_{\mathrm{DL}}[k] + \mathbf{n}_{\mathrm{DL}}[k],
\end{equation}
where $\mathbf{S}[k]=\left[\mathbf{F}_{\mathrm{RF}}[1]\mathbf{s}[1,k],\ldots ,\mathbf{F}_{\mathrm{RF}}[G]\mathbf{s}[G,k]\right]^{\rm T}\in \mathbb{C}^{G\times N_{\mathrm{AP}}}$ denotes the pilot signals transmitted from the AP, and $\mathbf{y}_{\mathrm{DL}}[k]=\left[y_{\mathrm{DL}}[1,k],\ldots,y_{\mathrm{DL}}[G,k]\right]^{\rm T}\in \mathbb{C}^{G\times 1}$ denotes the noisy received signal vector on the $k$th subcarrier after ResNet-DQ. Note that the effects of quantization and non-linear de-quantization on the channel output are all considered in the noise $\mathbf{n}_{\mathrm{DL}}[k]\! =\! \big[n_{\rm DL}[1,k],\ldots,n_{\rm DL}[G,k]\big]^{\rm T}\! \in\! \mathbb{C}^{G\times 1}$. As the procedure of ResNet-DQ is highly complex, we posit that the noise on different subcarriers $k$ follows an AWGN distribution with different power $\sigma_n^2[k]$, i.e., $\mathbf{n}_{\mathrm{DL}}[k]\! \sim\! \mathcal{CN} (\mathbf{0},\sigma_n^2[k]\mathbf{I})$. If there is no quantization and de-quantization process, $n_{\rm DL}[g,k]\! =\! \check{n}_{\mathrm{DL}}[g,k]$ and $y_{\mathrm{DL}}[1,k]\! =\! \check{y}_{\mathrm{DL}}[g,k]$, $\forall g,k$. {In order to harness the sparsity features of UM-MIMO channels illustrated in Section II-B, while simultaneously avoiding the prohibitive computational burden caused by an excessively high number of iterations in the conventional iterative channel estimation algorithms, we propose a generic knowledge-driven deep learning network. Specifically, under the beam squint effect, the conventional frequency-flat DFT dictionary widely used in massive MIMO systems would cause the severe virtual angle-domain support shift across different subcarriers. To solve this problem, we firstly propose two WRDs. The DFT-based WRD is tailored to the case of only far-field scatterers, and the data-driven WRD can be used for all cases including near-field scatterers. In this way, we can rewrite the channel estimation problem as a unified GMMV-CS problem, and the only difference is that the two dictionaries result in different measurement matrices. Secondly, we propose a generic GMMV-LAMP network by integrating trainable modules in the conventional GMMV-AMP algorithm to achieve acceleration and performance improvement.}

\subsubsection{DFT-based WRD}

The utilization of the DFT dictionary for sparse representation is effective when all scatterers are far-field for the AP as it uniformly divides the virtual angular domain into $N_{\mathrm{AP}}$ grids. However, there are two issues when the DFT dictionary is used directly across all subcarriers in wideband UM-MIMO systems. First, the virtual angular resolution of the DFT dictionary is limited to $N_{\mathrm{AP}}$, which may result in insufficient precision for off-grid scatterers. Second, the DFT dictionary works based on the assumption that the antenna spacing is approximately half of the wavelength for all subcarriers. However, when the bandwidth becomes ultra-wide and the beam squint effects appear, the frequency-flat DFT dictionary is not suitable. To address these issues, a frequency-dependent DFT-based WRD is proposed to accommodate the redundant virtual angle offsets on different subcarriers. In this way, \eqref{CE1} can be rewritten as

\begin{equation}
	\label{CE3} 
  \mathbf{y}_{\mathrm{DL}}[k] = \mathbf{S}[k] \mathbf{D}_{\mathrm{AD},\rho}[k] \mathbf{h}_{\mathrm{sparse}}[k] + \mathbf{n}_{\mathrm{DL}}[k],
\end{equation}
where $\mathbf{h}_{\mathrm{sparse}}[k]\in \mathbb{C}^{\rho N_{\mathrm{AP}}\times1}$ denotes the sparse channel vector in the virtual angular domain on the $k$th subcarrier. The DFT-based WRD {$\mathbf{D}_{\mathrm{AD},\rho}[k]\in \mathbb{C}^{N_{\mathrm{AP}}\times \rho N_{\mathrm{AP}}}$} is composed of several columns determined by the redundant factor $\rho$ and subcarrier index $k$, which can be written as

\begin{equation}
	\label{eqCEa} 
  \mathbf{D}_{\mathrm{AD},\rho}[k]\! =\! \left[\mathbf{a}_{0}[k],\mathbf{a}_{\frac{1}{\rho N_{\mathrm{AP}}}}\! \left[k\right],\ldots,\mathbf{a}_{\frac{\rho N_{\mathrm{AP}}\! -\! 1}{\rho N_{\mathrm{AP}}}}\! \left[k\right]\! \right]\!\! ,\!
\end{equation} 
where {$\mathbf{a}_{\frac{i}{\rho N_{\mathrm{AP}}}}\left[k\right]$} can be computed by \eqref{farfield-steeringvector}. In this way, the virtual angular domain is uniformly divided into $\rho N_{\mathrm{AP}}$ grids.
 
\begin{figure*}[tp!]
\vspace*{-4mm}
\begin{center}
\includegraphics[width=0.8\textwidth]{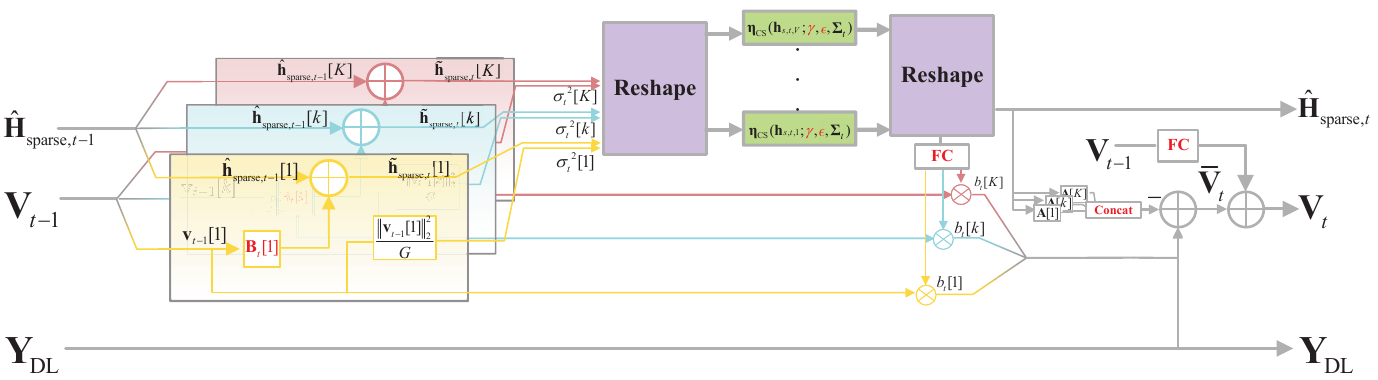}
\end{center}
\vspace*{-3mm}
\caption{The $t$th iteration/layer architecture of the proposed GMMV-LAMP network with trainable parameters.}
\label{LAMP-GMMV} 
\vspace*{-5mm}
\end{figure*}

\subsubsection{Data-Driven WRD}

Although the above DFT-based WRD fits the purely far-field scenario well, utilizing this DFT-based WRD in the hybrid near- and far- field scenario still leads to the degraded sparsity result. We further propose a data-driven frequency-dependent WRD to adaptively capture the sparsity characteristics of the near-field scattering propagation environment. Recall that the near-field array steering vectors \eqref{near-field vector} are defined by Cartesian coordinates, which can be transformed into the AoD-distance pairs. Specifically, given the Cartesian coordinate $(x_{l},y_{l})$ of the $l$th scatterer, the corresponding AoD-distance pair can be calculated as $(d_{l} ,\varphi_{l})$ where $d_{l}=\sqrt{x_{l}^2+y_{l}^{2}}$ and $\varphi_{l} = \arctan{(\frac{y_{l}}{x_{l}})}$. Correspondingly, $x_{l}$ and $y_{l}$ are given by $x_{l}=d_{l}\cos (\varphi_{l})$ and $y_{l}=d_{l}\sin (\varphi_{l})$. In light of this, 
we propose to introduce the trainable parameters $\mathbf{c}_{d}=\big[d_{0},d_{1},\ldots ,d_{V-1}\big]^{\rm T}\in\mathbb{R}^{V\times 1}$ and $\mathbf{c}_{\varphi}\! =\! \big[\varphi_{0},\varphi_{1},\ldots ,\varphi_{V-1}\big]^{\rm T}\! \in\! \mathbb{R}^{V\times 1}$ as the AoD-distance pairs to generate the corresponding columns in the data-driven WRD. The CE problem is then formulated as follows:
\begin{equation}\label{CE3.5} 
	\vspace*{-2.0mm}
\begin{array}{cl}
  \min\limits_{\mathbf{D}_{\mathrm{learn},(\mathbf{c}_{d},\mathbf{c}_{\varphi})}\left[k\right]} & \big\| \mathbf{h}_{\mathrm{sparse}}[k] \big\|_{0}, \\  
  \text{s.t.} & \mathbf{y}_{\mathrm{DL}}[k] = \mathbf{S}[k] \mathbf{D}_{\mathrm{learn},(\mathbf{c}_{d},\mathbf{c}_{\varphi})}[k] \mathbf{h}_{\mathrm{sparse}}[k]\\& \hspace{15mm}+ \mathbf{n}_{\mathrm{DL}}[k],
\end{array}
\end{equation}
where the data-driven WRD {$\mathbf{D}_{\mathrm{learn},(\mathbf{c}_{d},\mathbf{c}_{\varphi})}\left[k\right]$} is given by
{
\begin{align}\label{eqCEa1} 
	\vspace*{-2.0mm}
	& \mathbf{D}_{\mathrm{learn},(\mathbf{c}_{d},\mathbf{c}_{\varphi})}[k] = \big[\mathbf{a}_{(d_{0}\cos (\varphi_{0}),d_{0}\sin (\varphi_{0}))}[k],\ldots , \nonumber \\
	& \hspace*{14mm}\mathbf{a}_{(d_{V-1}\cos (\varphi_{V-1}),d_{V-1}\sin (\varphi_{V-1}))}[k]\big]\in \mathbb{C}^{N_{\mathrm{AP}}\times V} .
\end{align}
\vspace*{-2.0mm}
}

The $v$th column  $\mathbf{a}_{(d_{v-1}\cos (\varphi_{v-1}),d_{v-1}\sin (\varphi_{v-1}))}[k]$ is computed by \eqref{near-field vector}. It should be noted that as the degrees of freedom of the data-driven WRD increases, the number of columns $V$ necessarily increases to $V \gg \rho N_{\mathrm{AP}}$. 

\subsubsection{GMMV-LAMP}

In \eqref{CE1}, measurement matrices $\mathbf{A}[k]=\mathbf{S}[k]\mathbf{D}[k]$ vary with $k$ when the frequency-dependent pilot signals and WRDs are employed, where $\mathbf{D}[k]$ is either {$\mathbf{D}_{\mathrm{AD},\rho}[k]$ or $\mathbf{D}_{\mathrm{learn},(\mathbf{c}_{d},\mathbf{c}_{\varphi})}\left[k\right]$}. To take advantage of the exactly common support across all subcarriers, our GMMV-LAMP algorithm is proposed. The structure of the $t$th layer of this neural network is depicted in Fig.~\ref{LAMP-GMMV}, where $\mathbf{Y}_{\mathrm{DL}}\! =\!\left[\mathbf{y}_{\mathrm{DL}}[1],\ldots,\mathbf{y}_{\mathrm{DL}}[K]\right]\! \in\!\mathbb{C}^{G\times K}$, while $\hat{\mathbf{H}}_{\mathrm{sparse},t}\! =\! \left[\hat{\mathbf{h}}_{\mathrm{sparse},t}[1],\ldots,\hat{\mathbf{h}}_{\mathrm{sparse},t}[K]\right]\!\in\! \mathbb{C}^{V\times K}$ and $\mathbf{V}_{t}\! =\! \left[\mathbf{v}_{t}[1],\ldots,\mathbf{v}_{t}[K]\right]\! \in\! \mathbb{C}^{G\times K}$ denote the estimate and residual after the $t$th layer, respectively.

To describe our development of the GMMV-LAMP algorithm, we begin by offering an overview of the single measurement vector (SMV)-AMP algorithm. Specifically, we first ignore the common support feature and focus on an SMV CE problem on the $k$th subcarrier, which is formulated as 
\begin{equation}\label{CE4} 
\begin{array}{cl}
  \min\limits_{\mathbf{h}_{\mathrm{sparse}}[k]} & \big\| \mathbf{h}_{\mathrm{sparse}}[k] \big\|_{0}, \\  
  \text{s.t.} & \mathbf{y}[k] = \mathbf{A}[k] \mathbf{h}_{\mathrm{sparse}}[k] + \mathbf{n}[k] .
\end{array}
\end{equation}
With the initialization $\hat{\mathbf{h}}_{\mathrm{sparse},0}[k]\! =\! \mathbf{0}$ and $\mathbf{v}_{0}[k]\! =\! \mathbf{y}_{\rm DL}[k]$, $\forall k$, the $t$th iteration of SMV-AMP for $t=1,2,\ldots$ is as follows:
\begin{align}
	\tilde{\mathbf{h}}_{\mathrm{sparse},t}[k] =& \hat{\mathbf{h}}_{\mathrm{sparse},t-1}[k] + \mathbf{A}^{\rm H}[k] \mathbf{v}_{t-1}[k], \label{CE6.9} \\
	\sigma_{t}^{2}[k] =& \frac{1}{G}\left\|\mathbf{v}_{t-1}[k]\right\|_{2}^{2},	\label{CE7} \\
	\hat{\mathbf{h}}_{\mathrm{sparse},t}[k] =& \bm{\eta}\big(\tilde{\mathbf{h}}_{\mathrm{sparse},t}[k];\bm{\theta}_{t},\sigma_{t}[k]\big), \label{CE6} \\
	b_{t}[k] =& \frac{1}{G} \sum_{v=1}^{V} \frac{\partial\left[\bm{\eta} \left(\tilde{\mathbf{h}}_{\mathrm{sparse},t}[k];\bm{\theta}_{t},\sigma_{t}[k]\right)\right]_{v}}{\partial [\tilde{\mathbf{h}}_{\mathrm{sparse},t}[k]]_{v}},	\label{CE8} \\
  \mathbf{v}_{t}[k] =& \mathbf{y}[k] - \mathbf{A}[k] \hat{\mathbf{h}}_{\mathrm{sparse},t}[k] + b_{t}[k] \mathbf{v}_{t-1}[k] , \label{CE5} 
\end{align}
where $\bm{\eta}(\cdot;\bm{\theta}_{t},\sigma_{t}[k])$ is the shrinkage function with the parameter set $\bm{\theta}_t$ that is designed according to the chosen shrinkage function\footnote{In the common AMP-$l_{1}$ algorithm, $\bm{\eta}\left(\cdot;\bm{\theta}_{t},\sigma_{t}[k]\right)$ represents the `soft-thresholding' shrinkage and $\bm{\theta}_{t}=\alpha$ is a preset constant value \cite{AMP-L1}. But $\bm{\theta}_{t}$ can also be designed as layer-aware according to different shrinkage functions \cite{7934066}.}. The SMV-AMP algorithm consists of two main operations, the shrinkage function of \eqref{CE6} and `Onsager correction' of \eqref{CE5}. With the aid of the Onsager correction term $b_{t}[k]\mathbf{v}_{t-1}[k]$ in \eqref{CE5}, the input to the shrinkage function can be modeled as an AWGN-corrupted signal vector\footnote{In the $1$st iteration, $\tilde{\mathbf{h}}_{\mathrm{sparse},1}[k]$ can also be regarded as corrupted by AWGN, as we can imagine that there exists the $0$th iteration with $\mathbf{v}_{-1}[k]=\mathbf{0}$.} In addition, \eqref{CE6} achieves denoising based on the prior assumption of the noiseless signal.

Since there exists the common support characteristic across all the subcarriers, conventional AMP-type algorithms require a large number of iterations to converge. We propose the GMMV-LAMP network, which is summarized in Algorithm~\ref{Alg1}, to utilize the common support across all the subcarriers in a knowledge-driven DL manner. Compared with the conventional GMMV-AMP algorithm which takes tens or even hundreds of iterations to converge, our GMMV-LAMP utilizes a very small number of $T$ iterations/layers\footnote{This will be demonstrated in the simulation results section. to achieve CE by combining neural networks and expert knowledge of transmission functions.} We now explain our GMMV-LAMP algorithm in detail.

From Fig.~\ref{LAMP-GMMV}, it can be seen that there are three inputs to the $t$th layer of the GMMV-LAMP: the estimated sparse channel matrix $\hat{\mathbf{H}}_{\mathrm{sparse},t-1}$ and the updated residual $\mathbf{V}_{t-1}$ from the previous layer, and the received signal $\mathbf{Y}_{\rm DL}$ which is common to all the layers. We initialize	$\hat{\mathbf{H}}_{\mathrm{sparse},0}\! =\! \mathbf{0}$ and $\mathbf{V}_{0}\! =\! \mathbf{Y}_{\rm DL}$ in line~1, which is similar to the SMV-AMP algorithm. In the $t$th layer for $1\le t \le T$, we design a shrinkage function to leverage the exactly common support feature across all the subcarriers. Specifically, for each subcarrier, the corresponding $  \tilde{\mathbf{h}}_{\mathrm{sparse},t}[k] = \left[\tilde{h}_{1,t}[k],\ldots,\tilde{h}_{V,t}[k]\right]^{\rm T}\in\mathbb{C}^{V\times 1}$ is computed in line 4. Motivated by \cite{7934066}, we introduce a trainable matrix set $\mathbf{B}_{t}[k]\in\mathbb{C}^{G\times V}$, $\forall k\! \in\! \{1,\ldots,K\}$ and $\forall t\! \in\! \{1,\ldots,T\}$ in which $\mathbf{B}_{t}[k],\forall k$ are initialized as $\mathbf{A}[k]$ for all layers at the beginning of training.

\renewcommand{\algorithmicrequire}{\textbf{Input:}}
\renewcommand{\algorithmicensure}{\textbf{Output:}}
\begin{algorithm}[t]
	
\caption{Proposed GMMV-LAMP algorithm}
\label{Alg1}
\begin{algorithmic}[1]
	\small
	\REQUIRE De-quantized signal $\mathbf{Y}_{\mathrm{DL}}$, measurement matrices $\mathbf{A}[k]$, $\forall k\in\{1,2,...,K\}$, number of layers $T$;
	\ENSURE Estimated sparse matrix $\hat{\mathbf{H}}_{\mathrm{sparse},T}$;
	\STATE Initialize $\hat{\mathbf{H}}_{\mathrm{sparse},0} = \mathbf{0}$ and $\mathbf{V}_{0} = \mathbf{Y}_{\rm DL}$;
	\FOR {$t=1$ to $T$}
		\FOR {$k=1$ to $K$}			
			\STATE $\tilde{\mathbf{h}}_{\mathrm{sparse},t}[k]\leftarrow \hat{\mathbf{h}}_{\mathrm{sparse},t-1}[k]+\mathbf{B}_{t}^{\rm H}[k]\mathbf{v}_{t-1}[k]$;
			\STATE $\sigma_{t}^{2}[k]\leftarrow \frac{1}{G}\left\|\mathbf{v}_{t-1}[k]\right\|_{2}^{2}$;
		\ENDFOR
		\STATE $\mathbf{\Sigma}_{t} \leftarrow \mathrm{diag}\big(\sigma_{t}^{2}[1],\sigma_{t}^{2}[2],\ldots,\sigma_{t}^{2}[K]\big)$;
		\FOR {$v=1$ to $V$}
			\STATE $\tilde{\mathbf{h}}_{\mathrm{s},t,v}\leftarrow \Big[\tilde{h}_{v,t}[1],\ldots,\tilde{h}_{v,t}[K]\Big]^{\rm T}$;
			\STATE $\hat{\mathbf{h}}_{\mathrm{s},t,v}\leftarrow \bm{\eta}_{\rm CS}\left(\tilde{\mathbf{h}}_{\mathrm{s},t,v};\gamma,\epsilon,\mathbf{\Sigma}_{t}\right)$;
    \ENDFOR
		\FOR {$k=1$ to $K$}
		  \STATE $\hat{\mathbf{h}}_{\mathrm{sparse},t}[k]\leftarrow \left[\hat{h}_{1,t}[k],\ldots,\hat{h}_{V,t}[k]\right]^{\rm T}$;
		  \STATE $\bar{b}_{t}[k]\leftarrow \frac{1}{G} \sum_{v=1}^{V} \frac{\partial\left[\bm{\eta}_{\rm CS} \left(\tilde{\mathbf{h}}_{\mathrm{s},t,v};\gamma,\epsilon,\mathbf{\Sigma}_{t}\right)\right]_{k}}{\partial \left[\tilde{\mathbf{h}}_{\mathrm{s},t,v}\right]_{k}}$;
	  \ENDFOR
		\STATE $\bar{\mathbf{b}}_{t}\leftarrow \big[\bar{b}_{t}[1],\ldots,\bar{b}_{t}[K]\big]^{\rm T}$;
		\STATE $\mathbf{b}_{t}=\left[b_{t}[1],\ldots,b_{t}[K]\right]^{\rm T}\leftarrow g_{t}\left(\bar{\mathbf{b}}_{t}\right)$, where $g_{t}(\cdot)$ denotes an {\it FC};
		\FOR {$k=1$ to $K$}
			\STATE $\bar{\mathbf{v}}_{t}[k]\leftarrow\mathbf{y}[k] - \mathbf{A}[k] \hat{\mathbf{h}}_{\mathrm{sparse},t}[k] + b_{t}[k] \mathbf{v}_{t-1}[k]$;
		\ENDFOR
		\STATE $\bar{\mathbf{V}}_{t}\leftarrow\left[\bar{\mathbf{v}}_{t}[1],\ldots,\bar{\mathbf{v}}_{t}[K]\right]$;
	  \STATE $\mathbf{V}_{t}\leftarrow \bar{\mathbf{V}}_{t}+f_{t}\left (\mathbf{V}_{t-1}\right )$, where $f_{t}(\cdot)$ denotes an {\it FC}; 
		\STATE $\hat{\mathbf{H}}_{\mathrm{sparse},t}\leftarrow\left[\hat{\mathbf{h}}_{\mathrm{sparse},t}[1],\ldots,\hat{\mathbf{h}}_{\mathrm{sparse},t}[K]\right]$;
	\ENDFOR
	\STATE \textbf{Return} sparse matrix $\hat{\mathbf{H}}_{\mathrm{sparse},T}$.
\end{algorithmic}
\end{algorithm}
				
Inspired by the SMV-AMP algorithm, we view $\tilde{\mathbf{h}}_{\mathrm{sparse},t}[k]$ as an AWGN-corrupted sparse vector, and consider a denoising problem for the $v$th element of $\tilde{\mathbf{h}}_{\mathrm{sparse},t}[k]$, $\forall k$. In line~9, we collect the $v$th elements of $\tilde{\mathbf{h}}_{\mathrm{sparse},t}[k]$, $\forall k$, into $\tilde{\mathbf{h}}_{\mathrm{s},t,v} = \left[\tilde{h}_{v,t}[1],\ldots,\tilde{h}_{v,t}[K]\right]^{\rm T}\in\mathbb{C}^{K\times 1}$, which can be modeled as
\vspace*{-3.0mm}
\begin{equation}\label{CE11} 
	\vspace*{-2.0mm}
	\tilde{\mathbf{h}}_{\mathrm{s},t,v} = \bar{\mathbf{h}}_{\mathrm{s},v} + \mathbf{\Sigma}_{t}^{\frac{1}{2}}\mathbf{n}_{t,v},
\end{equation}
where $\bar{\mathbf{h}}_{\mathrm{s},v}\! =\! \left[\bar{h}_{v}[1],\ldots,\bar{h}_{v}[K]\right]^{\rm T}\! \in\! \mathbb{C}^{K\times 1}$ denotes the noiseless sparse vector, i.e., $\bar{h}_{v}[k]$ is the noiseless version of $\tilde{h}_{v,t}[k]$, and $\mathbf{n}_{t,v}\! \in\! \mathbb{C}^{K\times 1}\! \sim\! \mathcal{CN} (\mathbf{0},\mathbf{I})$. As aforementioned, owing to the nonlinear quantization and de-quantization process in obtaining $\mathbf{Y}_{\rm DL}$, the noise power on different subcarriers should be different. In this way, we use the diagonal matrix $\mathbf{\Sigma}_{t}\! =\! \mathrm{diag}\big(\sigma_{t}^{2}[1],\sigma_{t}^{2}[2],\ldots,\sigma_{t}^{2}[K]\big)\! \in\! \mathbb{C}^{K\times K}$ to represent the equivalent frequency-dependent noise power, where $\sigma_{t}^{2}[k]$ are computed in line 5, and typically $\sigma_{t}^{2}[i]\neq \sigma_{t}^{2}[j]$, $\forall i,j \in \{1,\ldots,K\}$ and $i\neq j$.

To recover $\bar{\mathbf{h}}_{\mathrm{s},v}$ from $\tilde{\mathbf{h}}_{\mathrm{s},t,v}$, we design a minimum mean square error (MMSE)-based denoiser. According to the common support property, $\bar{h}_{v}[k]$ occupies the identical sparsity for arbitrary subcarrier index $k$. Assume that $\bar{\mathbf{h}}_{\mathrm{s},v}$ follows the Bernoulli-Gaussian distribution, the prior of $\bar{\mathbf{h}}_{\mathrm{s},v}$ can be modeled as
\vspace*{-2.0mm}
\begin{equation}\label{CE13} 
	\vspace*{-2.0mm}
	p\big(\bar{\mathbf{h}}_{\mathrm{s},v};\gamma,\epsilon \big) =\left\{ \begin{array}{cl}
		1 - \gamma , & \text{if } \bar{\mathbf{h}}_{\mathrm{s},v}  =\mathbf{0} ,\\
		\gamma , & \text{if } \bar{\mathbf{h}}_{\mathrm{s},v}\sim \mathcal{CN}(\mathbf{0},\epsilon\mathbf{I}) ,
	\end{array} \right.
\end{equation}
where $\gamma$ is the probability that $\bar{h}_{v}[k] \neq 0$, $\forall k$, and $\epsilon$ is the variance of $\bar{h}_{v}[k]$ when $\bar{h}_{v}[k] \neq 0$, $\forall k$. The shrinkage function based on the MMSE denoiser $\bm{\eta}_{\rm CS}\left(\tilde{\mathbf{h}}_{\mathrm{s},t,v};\gamma,\epsilon,\mathbf{\Sigma}_{t}\right):\mathbb{C}^{K\times 1}\rightarrow\mathbb{C}^{K\times 1}$ is given by
\vspace*{-3.0mm}
\begin{align}\label{CE14} 
	\vspace*{-2.0mm}
	\hat{\mathbf{h}}_{\mathrm{s},t,v} &= \left[\hat{h}_{v,t}[1],\ldots,\hat{h}_{v,t}[k]\right]^{\rm T} = \bm{\eta}_{\rm CS}\left(\tilde{\mathbf{h}}_{\mathrm{s},t,v};\gamma,\epsilon,\mathbf{\Sigma}_{t}\right) \\&=\phi\big(\tilde{\mathbf{h}}_{\mathrm{s},t,v}\big) \mathrm{diag}\Big(\frac{\epsilon }{\epsilon+{\sigma_{t}^{2}[1]}},\ldots,\frac{\epsilon }{\epsilon+{\sigma_{t}^{2}[K]}}\Big)\tilde{\mathbf{h}}_{\mathrm{s},t,v} ,\!
\end{align}
{where $\phi(\tilde{\mathbf{h}}_{\mathrm{s},t,v})$ is detailed in Appendix based on \cite{shrinkage_function}.} In line~10, we compute $\hat{\mathbf{h}}_{\mathrm{s},t,v}\in\mathbb{C}^{K\times 1}$ for $1\le v\le V$. It should be noted that $\gamma$ and $\epsilon$ are independent of the indices $k$ and $v$ due to the random distribution of scatterers. Besides, in a relatively stable environment, $\gamma$ and $\epsilon$ tend to remain invariant for a long period, i.e., these two hyper-parameters can be regarded as independent of $t$. Therefore, $\gamma$ and $\epsilon$ are set as global trainable parameters, and they are learned from the existing datasets. It should be noted that we initialize $\gamma$ and $\epsilon$ to $\gamma_{0}=0$ and $\epsilon_{0}=1$ at the beginning of training, and all the layers utilize the same $\gamma$ and $\epsilon$ in the shrinkage function $\bm{\eta}_{\rm CS}\left(\tilde{\mathbf{h}}_{\mathrm{s},t,v};\gamma,\epsilon,\mathbf{\Sigma}_{t}\right)$, $\forall t$. After computing $\hat{\mathbf{h}}_{\mathrm{s},t,v}$, $\forall v$, we reshape all the vectors to obtain the updated sparse channel estimate $\hat{\mathbf{h}}_{\mathrm{sparse},t}[k]=\left[\hat{h}_{1,t}[k],\ldots,\hat{h}_{V,t}[k]\right]^{\rm T}$, $\forall k$, in line~13.

\begin{figure}[b!]
	\vspace*{-5mm}
	\begin{center}
		\includegraphics[width=0.5\textwidth]{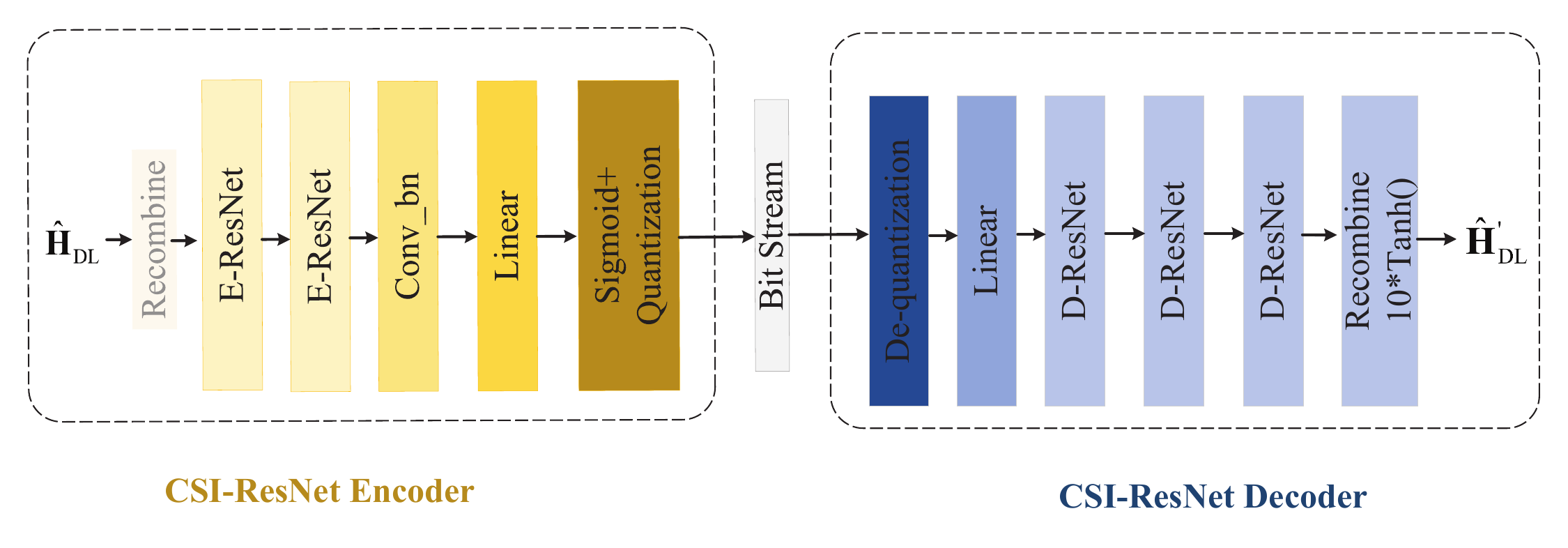}
	\end{center}
	\vspace*{-5mm}
	\caption{Structure of CSI-ResNet: a CSI-ResNet encoder compresses CSI to a bit vector and a CSI-ResNet decoder reconstructs the CSI matrix.}
	\label{AE-structure} 

\end{figure}
In line 14, lines 16 and 17, we employ DL networks to compute $\mathbf{b}_{t}=\left[b_{t}[1],\ldots,b_{t}[K]\right]^{\rm T}$. Specifically, $\bar{b}_{t}[k]$ for all the subcarriers are first calculated by utilizing the partial derivatives of the shrinkage function in line 14, which is similar to (\ref{CE8}) in the SMV-AMP, and they are collected together as $\bar{\mathbf{b}}_{t}=\big[\bar{b}_{t}[1],\ldots,\bar{b}_{t}[K]\big]^{\rm T}$ in line 16. In line 17, $\bar{\mathbf{b}}_{t}$ is utilized as the input of an FC layer $g_{t}(\cdot):\mathbb{C}^{K\times 1}\rightarrow\mathbb{C}^{K\times 1}$ to obtain a refined result $\mathbf{b}_{t}=g_{t}\left(\bar{\mathbf{b}}_{t}\right)$. 
\begin{figure}[!t]

	\begin{center}
		\includegraphics[width=0.5\textwidth]{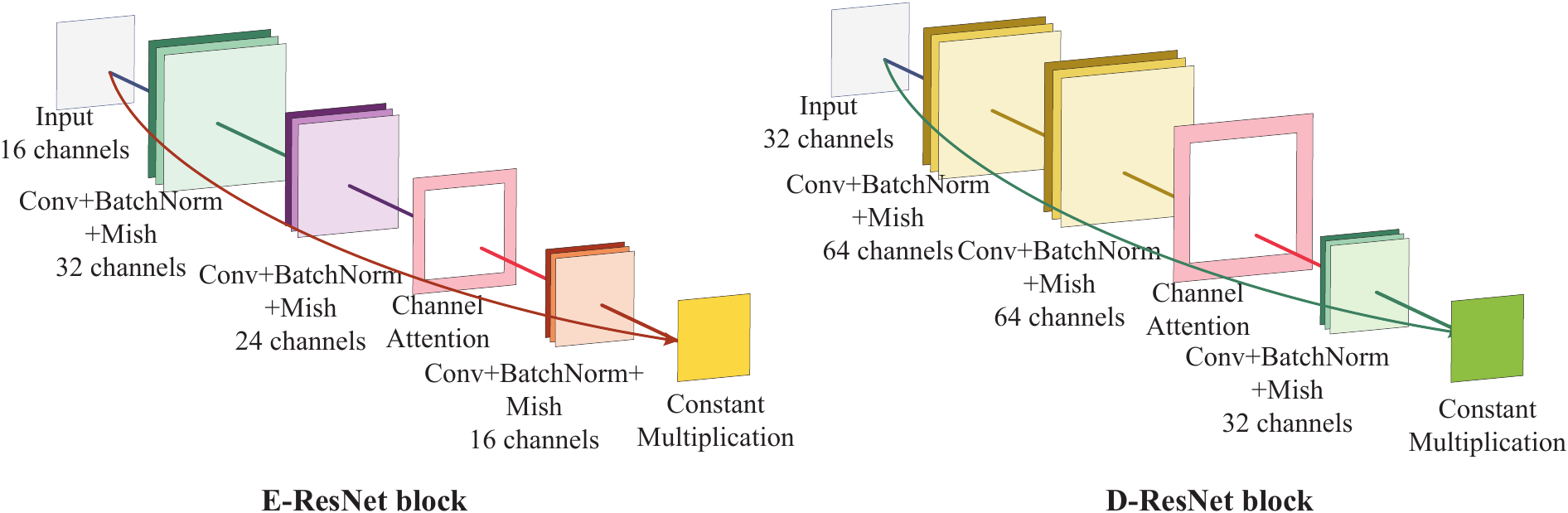}
	\end{center}
	\vspace*{-5mm}
	\caption{Structures of E-ResNet block and D-ResNet block.}
		\vspace*{-5mm}
	\label{resnet} 

\end{figure}
In line 19, we update a coarse residual $\bar{\mathbf{v}}_{t}[k]$ with the aid of the Onsager correction term $b_{t}[k]\mathbf{v}_{t-1}[k]$ for each subcarrier, which is the same as \eqref{CE5} in the SMV-AMP. Here, inspired by the concept of momentum in DL, we utilize the residual $\mathbf{V}_{t-1}$ from the $(t-1)$th layer to assist in accelerating convergence in the $t$th layer. Specifically, we collect $\bar{\mathbf{v}}_{t}[k]$, $\forall k$, into $\bar{\mathbf{V}}_{t}\! =\! \left[\bar{\mathbf{v}}_{t}[1],\ldots,\bar{\mathbf{v}}_{t}[K]\right]\! \in\! \mathbb{C}^{G\times K}$, and utilize an FC layer $f_{t}(\cdot):\mathbb{C}^{G\times K}\! \rightarrow\! \mathbb{C}^{G\times K}$ to obtain $\mathbf{V}_{t}\! =\! \left[\mathbf{v}_{t}[1],\ldots,\mathbf{v}_{t}[K]\right]\! =\! \bar{\mathbf{V}}_{t}+f_{t}\left (\mathbf{V}_{t-1}\right)$.

After $T$ iterations/layers, the GMMV-LAMP network outputs the final estimated sparse matrix $\hat{\mathbf{H}}_{\mathrm{sparse},T}\! =\!\left[\hat{\mathbf{h}}_{\mathrm{sparse},T}[1],\ldots,\hat{\mathbf{h}}_{\mathrm{sparse},T}[K]\right]\! \in\! \mathbb{C}^{V\times K}$, and the CSI in the spatial-frequency domain can be computed as $\hat{\mathbf{H}}_{\mathrm{DL}}\! =\! \left[\mathbf{D}[1]\hat{\mathbf{h}}_{\mathrm{sparse},T}[1],\ldots,\mathbf{D}[K]\hat{\mathbf{h}}_{\mathrm{sparse},T}[K]\right]\! \in\! \mathbb{C}^{N_{\rm AP}\times K}$.
\vspace*{-4.0mm}
\subsection{Data-Driven CSI Feedback Module}\label{S3.3}

After the UE has estimated $\hat{\mathbf{H}}_{\mathrm{DL}}$, the CSI matrix is fed back to the AP for downlink beamforming. However, if the CSI matrix is fed back to the AP without compression, the feedback overhead is unaffordable. For example, assume that a system with $N_{\mathrm{AP}}=64$ and $K=64$ is achieving CSI feedback and 32-bits accuracy of floating-point numbers is considered. Then, the total feedback overhead are $2\times 64\times64\times32 = 264144$\,bits, where the factor 2 is owing to the fact that each channel coefficient has real and imaginary parts. One feasible CSI feedback method is to compress the floating-point numbers into low-bit types, such as $3$ or $2$ bits, but this results in significant quantization noise.

The autoencoder is a widely utilized deep neural network structure \cite{DLdetection2,Tranf} to perform feature extraction for CSI feedback \cite{8322184,8482358}. The autoencoder is designed to be an end-to-end system, with the goal of reconstructing a noiseless version of the target. In this paper, we propose a data-driven autoencoder called CSI-ResNet which consists of a CSI-ResNet encoder and a CSI-ResNet decoder as depicted in Fig.~\ref{AE-structure}.

\subsubsection{CSI-ResNet Encoder}

The estimated CSI $\hat{\mathbf{H}}_{\mathrm{DL}}$ is divided into $\Re\big(\hat{\mathbf{H}}_{\mathrm{DL}}\big)$ and $\Im\big(\hat{\mathbf{H}}_{\mathrm{DL}}\big)$, and then stacked as the input to the CSI-ResNet encoder. Two `E-ResNet' blocks are utilized for feature extraction. The structure of E-ResNet, as illustrated in Fig.~\ref{resnet}, is inspired by ResNet-DQ but with a higher kernel size and more channels. It includes three subblocks, each consisting of a `Conv' layer, a `BatchNorm' layer, and a `Mish' activation layer. The first subblock increases the number of channels from 16 to 32, and the second subblock reduces the number of channels to 24. Then a `Channel Attention' layer is used to find the correlation across different channels. In the third subblock, the input is convolved to generate a 16-channel output for the final skip connection. After the two E-ResNet blocks, a `Convbn' block which includes a `Conv' layer and a `BatchNorm' layer performs further feature extraction. Although most of the information has been retained, the dimension of output is reduced for CSI compression. The `Linear' block then compresses the CSI as a floating-point vector. Finally, the floating-point vector is quantized into a bit vector using the `Sigmoid+Quantization' block. Specifically, the sigmoid function limits the value range of the floating-point number to $[0, \, 1]$, and the floating-point vector is quantized into a bit stream. For example, if the number of quantization bits is set to $2$, the output range of each element in the floating-point vector is divided into $[0,\,0.25)$, $[0.25,\,0.5)$, $[0.5,\,0.75)$ and $[0.75,\,1]$, and the corresponding bit vectors are `$00$', `$01$', `$10$' and `$11$'.

\subsubsection{CSI-ResNet Decoder}

The `De-quantization' block transforms the bit vector into a floating-point vector with 32 channels. Subsequently, a `Linear' block and three `D-ResNet' blocks are utilized. The D-ResNet block, shown in Fig.~\ref{resnet}, has a similar structure to the E-ResNet block but with different dimensions in its subblocks. The output after the three D-ResNet blocks is real-valued and contains both real and imaginary parts. With the aid of the `Tanh' activation layer and an amplitude adjustment operation, the $2$-channel real-valued tensor is then recast as the complex-valued CSI matrix $\hat{\mathbf{H}}'_{\mathrm{DL}}$.
\vspace*{-5.0mm}
\subsection{Training Strategy}\label{S3.4}

Due to large fluctuations in the multipath component gains, we consider the normalized mean square error (NMSE) as the cost function, which is written as
\begin{equation}\label{NMSE} 
	Loss = \mathrm{NMSE}\big(\hat{\mathbf{X}};\mathbf{X}\big)=\mathbb{E}\left [\frac{\left\| \hat{\mathbf{X}}-\mathbf{X} \right \|_{F}^{2}}{\left \| \mathbf{X} \right \|_{F}^{2}}\right ],
\end{equation}
where $\hat{\mathbf{X}}$ and $\mathbf{X}$ denote the estimated signal and the true signal, respectively. There are three distinct modules in the whole network. Specifically, ResNet-DQ is data-driven, GMMV-LAMP is knowledge-driven, and CSI-ResNet is data-driven, and they have different designs and training targets. Therefore, an end-to-end training approach may not be appropriate. Therefore, we apply separate training procedures for the ResNet-DQ, GMMV-LAMP and CSI-ResNet modules, and train them in sequence with different loss functions. 

First, the loss function of ResNet-DQ is the NMSE between the output $\mathbf{Y}_{\mathrm{DL}}\! =\! \left[\mathbf{y}_{\mathrm{DL}}[1],\ldots,\mathbf{y}_{\mathrm{DL}}[K]\right]$ and the infinite-resolution noiseless signal $\bar{\mathbf{Y}}_{\mathrm{DL}}\! =\! \big[\bar{\mathbf{y}}_{\mathrm{DL}}[1],\ldots,\bar{\mathbf{y}}_{\mathrm{DL}}[K]\big]$ in the frequency domain, i.e.,
\begin{equation}\label{cost_ResNet-DQ} 
	Loss_{\mathrm{ResNet-DQ}} = \mathrm{NMSE}\big(\mathbf{Y}_{\mathrm{DL}};\bar{\mathbf{Y}}_{\mathrm{DL}}\big).
\end{equation}

After ResNet-DQ is trained, it is set to the evaluation mode, and the GMMV-LAMP network is trained with the cost function chosen to be the NMSE between the estimated CSI $\hat{\mathbf{H}}_{\mathrm{DL}}$ and the perfect CSI $\bar{\mathbf{H}}_{\mathrm{DL}}$, namely, 
\begin{equation}\label{cost_GMMV-LAMP} 
	Loss_{\mathrm{GMMV-LAMP}} = \mathrm{NMSE}\big(\hat{\mathbf{H}}_{\mathrm{DL}};\bar{\mathbf{H}}_{\mathrm{DL}}\big).
\end{equation}

The parameters in our GMMV-LAMP network are divided into global parameters, such as $\gamma, \epsilon$, and layer-level parameters, such as $\mathbf{B}_t[k]$. The training of the GMMV-LAMP network is in an all-layer manner. Specifically, when the $t$th layer is trained, the layer-level parameters in all $t'$th $\left(1\le t'\le t \right)$ layers and global parameters are trained. Note that when the data-driven WRD is trained, $\mathcal{C}_{d}$ and $\mathcal{C}_{\varphi}$ are only activated for training the first and second layers. Once the second layer has been trained, $\mathcal{C}_{d}$ and $\mathcal{C}_{\varphi}$ remain fixed.

After the GMMV-LAMP is trained, we generate the estimated CSI in the training and validation sets based on the trained ResNet-DQ and GMMV-LAMP for CSI feedback. For the CSI-ResNet, which is composed of both the encoder and decoder, an end-to-end training strategy is employed, and the cost function is defined as the NMSE between the reconstructed CSI $\hat{\mathbf{H}}'_{\mathrm{DL}}$ and the perfect CSI $\bar{\mathbf{H}}_{\mathrm{DL}}$, i.e.,
\begin{equation}\label{cost_CSI-ResNet} 
	Loss_{\mathrm{CSI-ResNet}} = \mathrm{NMSE}\left(\hat{\mathbf{H}}'_{\mathrm{DL}};\bar{\mathbf{H}}_{\mathrm{DL}}\right).
\end{equation}

\section{Simulation Results}\label{S4}
\begin{figure*}[t!]
		\vspace*{-5mm}
	\begin{center}
		\includegraphics[width=0.7\textwidth]{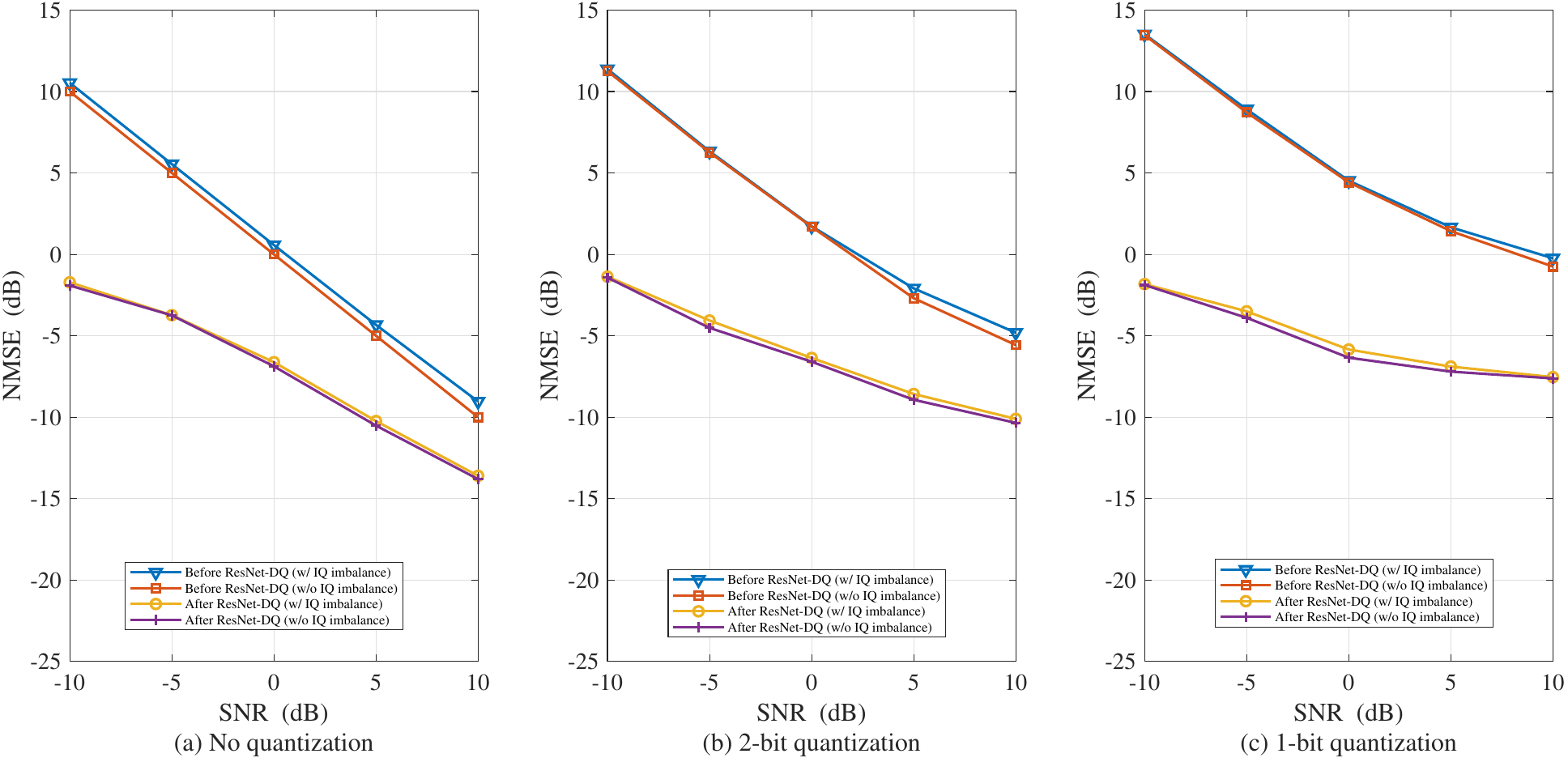}
	\end{center}
	\vspace*{-5mm}
	\caption{De-quantization NMSE with different resolutions of ADCs: (a) infinite-resolution ADCs, (b) $2$-bit ADCs, (c) $1$-bit ADCs.}
	\label{dequan_result} 
	\vspace*{-6mm}
\end{figure*}

This section provides simulation results to verify the effectiveness of the proposed approach in a typical mmWave wideband UM-MIMO system. There are $N_{\mathrm{AP}}=128,\, 256$ and $512$ antennas at the AP with $N_{\mathrm{RF}}=2$ RF chains\footnote{The Rayleigh distance of the 128-element and 256-element ULAs in such a parameter setup is respectively around 8.8 meters and 35 meters, whose array aperture can be regarded to be extra-large for most indoor communication scenarios \cite{UMMIMO_antenna}.}, and the number of pilot symbols is set as $G=32,\, 64$. The IQ gain and phase error factors are set as $\zeta _{A}=0.1$ $\zeta _{\theta}=5^{\circ}$, respectively. The carrier frequency is 70\,GHz, the bandwidth is 10\,GHz while there are $K=64$ subcarriers at the CE stage. For purely far-field, purely near-field, and hybrid near- and far- field scenarios, we generate different datasets for training. Each sample of the aforementioned datasets is composed of several near-field or far-field or hybrid multipath components that are generated with different policies. Specifically, each far-field path is generated with a random AoD within $[-\frac{\pi}{3},\frac{\pi}{3}]$ and path delay within $6.4\times10^{-9}$ seconds. Based on the effective Rayleigh distance \cite{9685542}, we generate near-field paths within the effective Rayleigh distance area. More specifically, a random AoD is selected and the corresponding effective Rayleigh distance is calculated, and one distance is generated within the effective Rayleigh distance so that one AoD-distance pair can be obtained.

Unless stated otherwise, the number of scatterers in one sample is $L\! =\! 6$, and each CSI sample in the hybrid near- and far- field dataset is composed of 3 far-field scatterers and 3 near-field scatterers, while the CSI samples in the purely far-field dataset and the purely near-field dataset include 6 far-field scatterers and 6 near-field scatterers, respectively. In all simulations, the CSI samples in the training sets and the validation sets have identical features. Moreover, the number of antennas at the AP is $N_{\rm AP} = 128$, and the number of pilot symbols is $G=32$. In the rest figures, the x-axis represents the SNR level of AWGN which ranges over -10\,dB to 10\,dB, that is, the power ratio of the perfectly received signals to AWGN. For different modules, we choose the NMSE of different outputs as the metric for performance evaluation. Specifically, we utilize $\mathrm{NMSE}\big(\mathbf{Y}_{\mathrm{DL}};\bar{\mathbf{Y}}_{\mathrm{DL}}\big)$, $\mathrm{NMSE}\big(\hat{\mathbf{H}}_{\mathrm{DL}};\bar{\mathbf{H}}_{\mathrm{DL}}\big)$ and $\mathrm{NMSE}\left(\hat{\mathbf{H}}'_{\mathrm{DL}};\bar{\mathbf{H}}_{\mathrm{DL}}\right)$ as the metrics of the de-quantization NMSE, estimation NMSE and reconstruction NMSE, respectively. The number of training samples and test samples are 150000 and 30000, respectively. In the training stage, there are 20, 10, and 80 epochs for Res-DQ, GMMV-LAMP, and CSI-ResNet, respectively. The batchsize is set as 180, and the Adam optimizer with a learning rate of 0.001 is employed. The experiments are performed in Visual Studio Code (Python 3.9.7 and pytorch 1.10) on a computer with Nvidia TITAN RTX.

\subsection{Data-Driven ResNet-DQ}\label{S4.1}
\begin{figure}[b!]
\vspace*{-5.5mm}	
	\centering
	\includegraphics[scale=0.48]{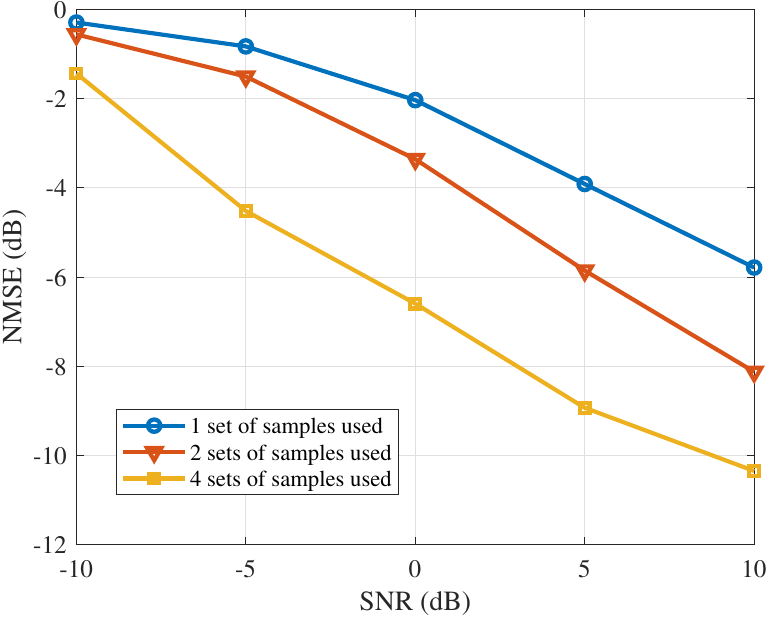}
	\vspace*{-1mm}
	\caption{De-quantization NMSE given different sets of time-domain oversampling samples.}
	\label{dequan_channel}
	
\end{figure}

Figure.~\ref{dequan_result} depicts the performance of ResNet-DQ as the function of SNR. The y-axis denotes the NMSE between the de-quantized signal matrix $\mathbf{Y}_{\mathrm{DL}}$ and the perfectly received frequency-domain signal $\bar{\mathbf{Y}}_{\mathrm{DL}}$, which is a measure of the equivalent noise in the received signal. When there is no quantization as shown in Fig.~\ref{dequan_result}\,(a), ResNet-DQ achieves the NMSE improvements of 10\,dB and 4\,dB at the SNRs of $-10$\,dB and 10\,dB, respectively. Compared to infinite-bit ADCs, the presence of quantization noise from 2-bit ADCs greatly impacts the quality of the output signal, especially at high SNRs. Specifically, there exists a large NMSE loss when no de-quantizer is utilized, even at the SNR of 10\,dB. However, once ResNet-DQ is employed, the NMSE can be reduced to -9\,dB at the SNR of 10\,dB, and moreover ResNet-DQ achieves the NMSE improvements of 13\,dB and 5\,dB at the SNRs of $-10$\,dB and 10\,dB, respectively, as shown in Fig.~\ref{dequan_result}\,(b). When 1-bit quantization is considered, as depicted in Fig.~\ref{dequan_result}\,(c), the NMSE can be reduced to -7\,dB at the SNR of 10\,dB with the aid of ResNet-DQ, and the NMSE improvements of 15\,dB and 7\,dB are achieved at the SNRs of $-10$\,dB and 10\,dB, respectively. The results of Fig.~\ref{dequan_result} clearly demonstrate the effectiveness of our ResNet-DQ. Additionally, in the presence of IQ imbalance, the neural network shows negligible performance loss in all the cases, which can be attributed to the robustness of our design to detect the correlation between the real and imaginary parts of the signal in the time domain. In the following simulations (e.g., Fig. 12-22, Table II, and Table III), the IQ imbalance and $2$-bit quantization are considered.

To verify the performance gain of utilizing multiple sets of time-domain oversampling samples, Fig.~\ref{dequan_channel} shows the de-quantization performance of ResNet-DQ given different numbers $W$ of input sets of samples when $2$-bit ADCs are employed. The results clearly demonstrate that as the number of input sets $W$ increases, the de-quantization effect becomes better and the NMSE decreases significantly. In particular, the NMSE reduction between the $W\! =\! 4$ input sets and the $W\! =\! 2$ input sets is considerably larger than that between the $W\! =\! 2$ input sets and the $W\! =\! 1$ input set. Considering this de-quantization performance, the number of input sets is set to $W\! =\! 4$ in the rest simulations.

\begin{figure}[b!]
	\vspace*{-5mm}	
	\centering
	\includegraphics[scale=0.32]{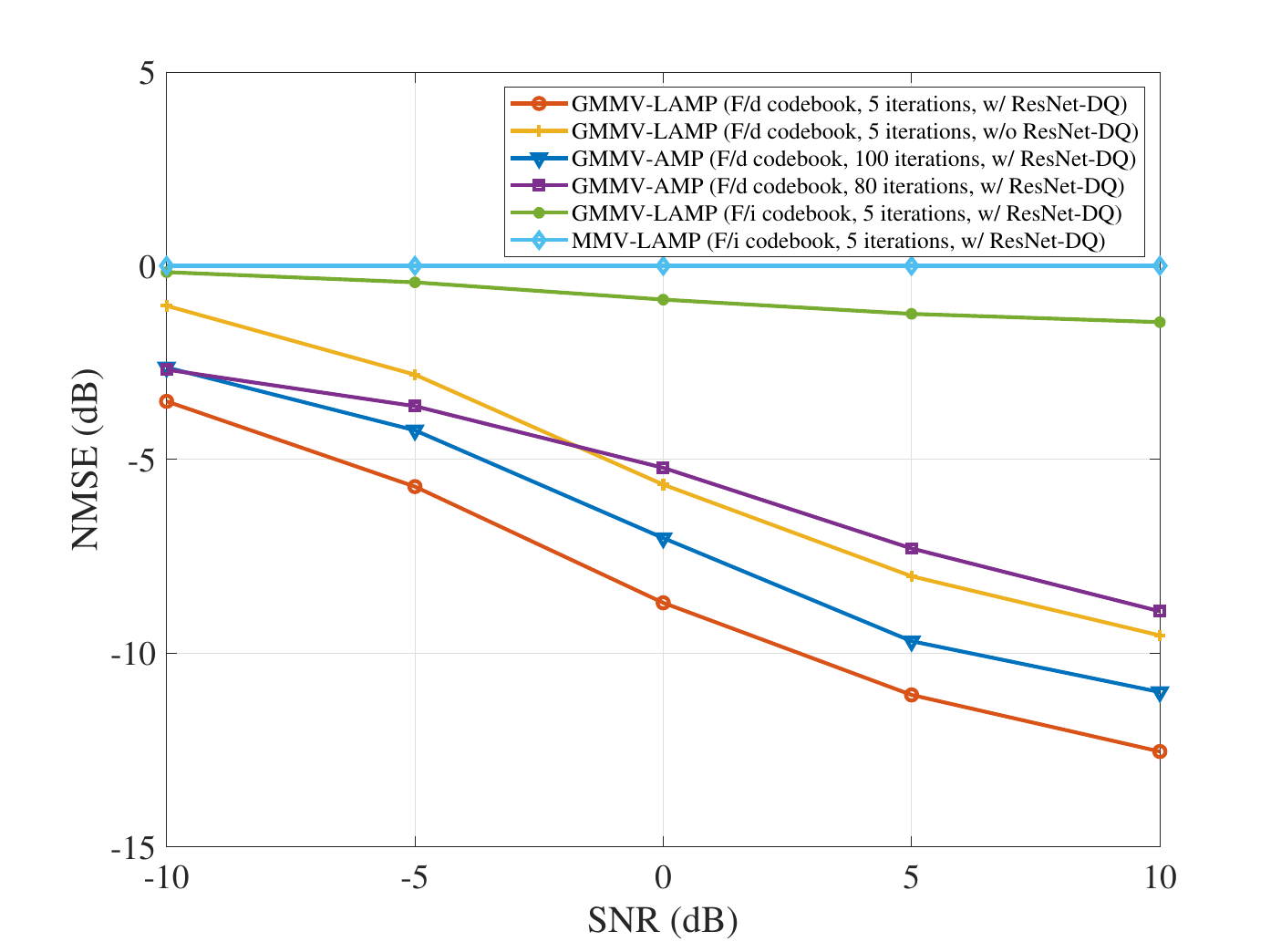}
	\vspace*{-3mm}	
	\caption{Comparison of the estimation NMSE for different CE schemes in the purely far-field scenario.}
	\label{LAMP_GMMV_CE}
	\vspace*{-1mm}	
\end{figure}

\subsection{Knowledge-Driven GMMV-LAMP}\label{S4.2}

\begin{figure}[t!]
		\vspace*{-5mm}	
	\centering
	\includegraphics[scale=0.36]{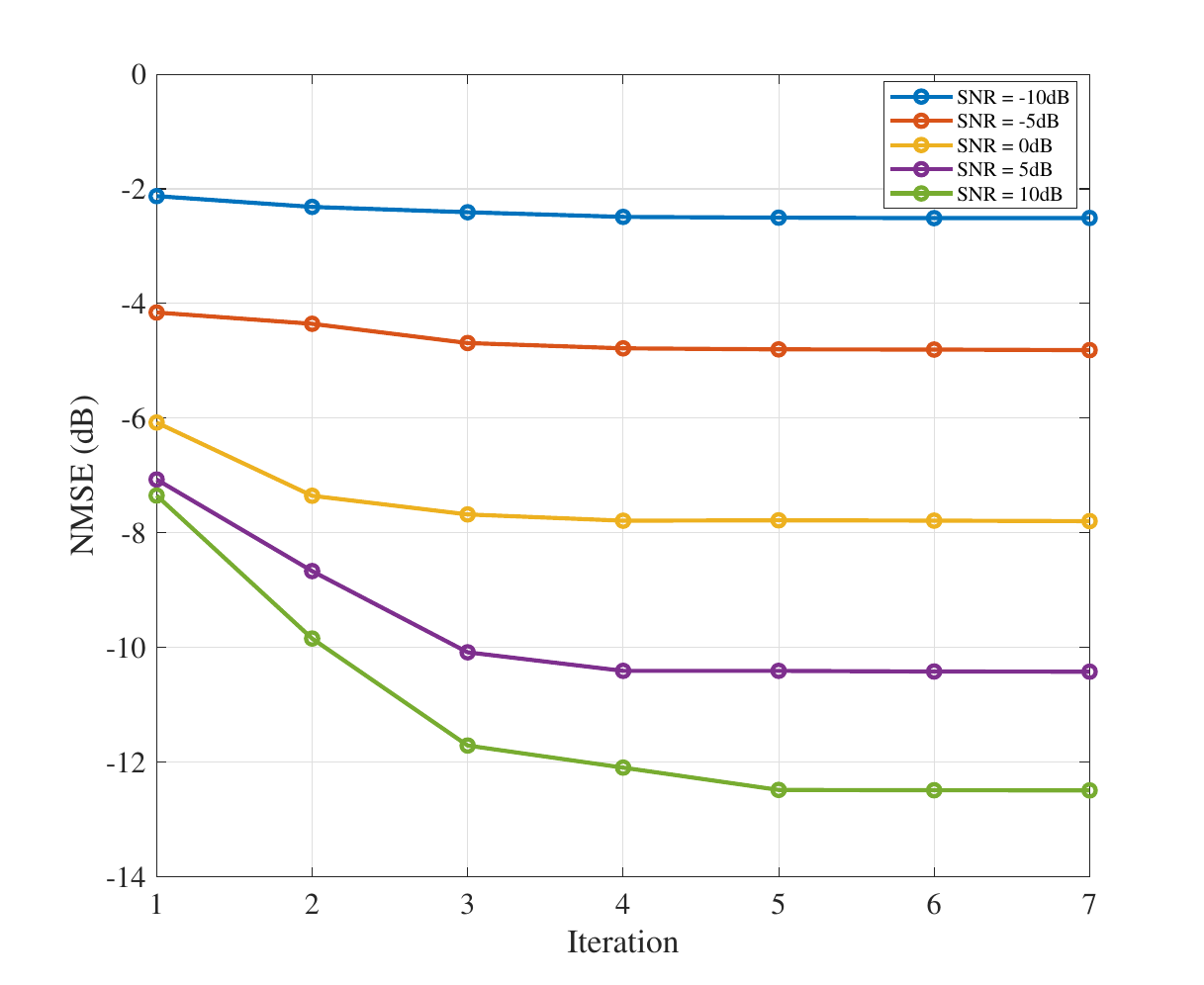}
	\vspace*{-3mm}
	\caption{Estimation NMSE as the function of number of iterations/layers given different SNRs.}
	\label{Different-SNR}
	\vspace*{-3mm}
\end{figure}
\subsubsection{Purely Far-Field Scenario}

The NMSE performance of different CE approaches is compared in Fig.~\ref{LAMP_GMMV_CE}. It should be mentioned that for all methods the redundant factor is set as $\rho=4$. The abbreviations 'F/d codebook' and 'F/i codebook' in the legend denote the frequency-dependent DFT WRD and the frequency-flat DFT WRD, respectively, while 'w/ ResNet-DQ' and 'w/o ResNet-DQ' represents that the input signal is the de-quantized signal by ResNet-DQ and the raw signal, respectively. The benchmark MMV-LAMP network \cite{9452036} assumes that the identical pilot matrices and frequency-flat DFT WRD are utilized, that is, the network is designed based on the same measurement matrices for all subcarriers. It can be seen that the MMV-LAMP network fails to work. As aforementioned, the GMMV-AMP \cite{8961111} suffers from slow convergence and has difficulty effectively coping with hardware imperfections, which is self-evident in the results of Fig.~\ref{LAMP_GMMV_CE}. {Clearly, the proposed GMMV-LAMP network (F/d codebook, 5-iterations, w/ ResNet-DQ) outperforms the above two candidate approaches, in terms of CE accuracy and convergence speed.} Additionally, the results of Fig.~\ref{LAMP_GMMV_CE} also confirm that it is important to utilize the frequency-dependent DFT WRD and perform ResNet-DQ in the GMMV-LAMP network.

With the identical system setup to Fig.~\ref{LAMP_GMMV_CE}, the convergence performance of the proposed GMMV-LAMP network with the frequency-dependent DFT WRD is investigated in Fig.~\ref{Different-SNR}. It is observed that the CE accuracy improves most in the first iteration/layer, and this can be attributed to the fact that the learnable parameters accelerate the convergence in the GMMV-LAMP network. It can also be seen that the GMMV-LAMP network converges after $T=5$ iterations. Since our GMMV-LAMP network can converge in just a few iterations, its time complexity is very low.

\begin{figure}[t!]
	
	\centering
	\includegraphics[scale=0.55]{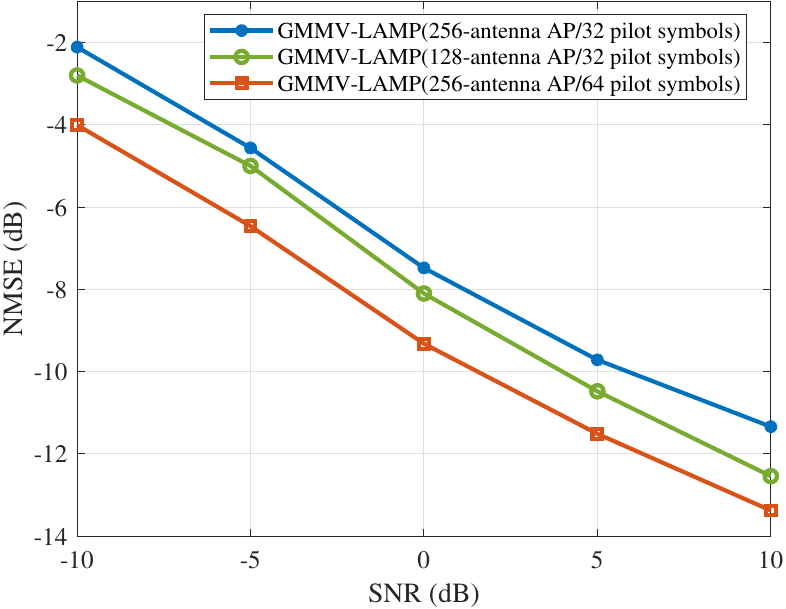}
	\vspace*{-4mm}
	\caption{Estimation NMSE as the function of the number of antennas at the AP $N_{\rm AP}$ and pilot symbols $G$.}
	\label{Different-antennas}
	\vspace*{-3mm}
\end{figure}

We next investigate the impacts of the number of pilot symbols $G$ and the number of AP antennas $N_{\mathrm{AP}}$ on the achievable CE performance in Fig.~\ref{Different-antennas} for our GMMV-LAMP, with the redundant factor of $\mathbf{D}_{\rm AD}$ fixed to $\rho\! =\! 4$. Given $G\! =\! 32$, the NMSE for the case of 128 AP antennas is slightly better than that of 256 AP antennas. This is due to the fact that although the dimension of CSI with 256 antennas is doubled compared to that with 128 antennas, the virtual angular sparsity also improves as $N_{\mathrm{AP}}$ increases. Additionally, it can be seen that by doubling the pilot symbols from 32 to 64, significant performance improvement can be achieved. This is because more pilot symbols provide more information with the same prior distribution. In the rest simulations, the number of antennas at the AP and the number of pilot symbols are set to $N_{\rm AP} = 128$ and $G=32$, respectively. 
{\renewcommand {\arraystretch}{1}
	\begin{table*}[b!]
		\vspace*{-2mm}
		
		\caption{Estimation NMSE (dB) versus Bandwidth, the Number of AP Antennas $N_{\mathrm{AP}}$ and the Number of Pilot Symbols $G$.} 
		\vspace*{-2mm}
		\label{major_add} 
		\begin{center}
			\setlength{\tabcolsep}{1mm}
			
			\begin{tabular}{c|c|c|c|c|c|c|c|c|c|c|c|c}
				
				\Xhline{1.2pt}
				\multirow{3}{*}{$N_{\mathrm{AP}}\,\&\, G$}              & \multicolumn{4}{c|}{1\,GHz}                         & \multicolumn{4}{c|}{5\,GHz}      & \multicolumn{4}{c}{10\,GHz}                   \\ \cline{2-13} 
				& \multicolumn{2}{c|}{Proposed} &\multirow{2}{*}{GMMV-AMP} & \multirow{2}{*}{SOMP} & \multicolumn{2}{c|}{Proposed} & \multirow{2}{*}{GMMV-AMP} & \multirow{2}{*}{SOMP}& \multicolumn{2}{c|}{Proposed} & \multirow{2}{*}{GMMV-AMP}&  \multirow{2}{*}{SOMP} \\ \cline{2-3} \cline{6-7} \cline{10-11} 
				& \multicolumn{1}{c|}{WRD} & \multicolumn{1}{c|}{Flat} & & & \multicolumn{1}{c|}{WRD} & \multicolumn{1}{c|}{Flat}&   && \multicolumn{1}{c|}{WRD} & \multicolumn{1}{c|}{Flat} &  &\\ 
				\Xhline{1.2pt}
				$N_{\mathrm{AP}}=128$, $G=32$   &     -12.7182      &-11.3321&-8.7002&-8.2498 &      -12.5793  &   -9.2145 &-8.5594&-3.7001 & -12.6653&-8.2210&-8.5961 &-2.0672 \\ \hline
				$N_{\mathrm{AP}}=256$, $G=32$   &     -11.2955      &-7.9770&-4.8413&-5.8990 &      -10.4080       & -2.4824 &-5.2741&-1.2284 & -11.3075&-0.9361 &-5.2918&-0.6148 \\\hline
				$N_{\mathrm{AP}}=256$, $G=64$   &     -13.3765     & -10.8847&-10.1446&-6.6207  &      -13.0043& -7.5848 &-9.9922&-1.9813 & -13.2604&-5.0731 &-10.1782& -1.0638 \\\hline
				$N_{\mathrm{AP}}=512$, $G=32$   &     -10.1422      &-1.6620&-5.5594&-2.2167 &       -10.0240  & -0.0261 &-5.7251&-0.1122 &  -10.1653&  -0.1053&-5.6037 &0.2998 \\\hline
				$N_{\mathrm{AP}}=512$, $G=64$   &     -12.1427       &-9.2477&-6.1474& -4.0434 &      -11.8233 &-1.5348&-6.0643& -1.0466 & -12.6653&-0.9088 &-6.0502 & -0.2407 \\
				\Xhline{1.2pt}
			\end{tabular}
		\end{center}
		
	\end{table*}

}

{In Table~\ref{major_add}, the influence of $N_{\mathrm{AP}}$ and the system bandwidth is systematically presented. Moreover, we also provide the results of simultaneous OMP (SOMP) mentioned in [17], where the pilot symbols are set to be the same across all subcarriers for convenience. We set the damping factor in GMMV-AMP with frequency-dependent WRDs to 0.9 to prevent the algorithm's divergence, and the number of iterations is set to 80. It should also be noted that the redundant factors $\rho$ are set to 4 for all candidate values of $N_{\mathrm{AP}}$s, and the training and testing SNRs are both set to 10\,dB. Firstly, we can observe that utilizing frequency-dependent WRDs guarantees the effectiveness of GMMV-LAMP under all settings compared with the frequency-flat dictionaries. It should be noted that the simulation results with frequency-dependent WRDs will be an essential assessment of the beam squint effect. We can find a discernible exacerbation of the beam squint effect concurrent with the escalations in both the system bandwidth and the number of antennas. On the one hand, by fixing the system bandwidth and other parameters, an increment in $N_{\mathrm{AP}}$ results in a larger gap between utilizing frequency-dependent WRDs and frequency-flat dictionaries in GMMV-LAMP. For example, when $N_{\mathrm{AP}}=128$, $G=32$ and the system bandwidth is 1\,GHz, the estimation NMSE gap between two kinds of dictionaries is around 1.4\,dB, while the gap reaches around 3.3\,dB when $N_{\mathrm{AP}}$ increases to 256. Even if the number of pilot symbols $G$ becomes 64, the estimation NMSE gap is still larger than 2\,dB. Furthermore, for almost all candidate values of $N_{\mathrm{AP}}$ and $G$, the estimation NMSE gap increases with the growth of bandwidth from 1\,GHz to 10\,GHz. Specifically, when $N_{\mathrm{AP}}=256$ and $G=64$, the estimation NMSE gap increases from 2.4\,dB to 8.2\,dB. However, the estimation NMSE gap is around 10\,dB when $N_{\mathrm{AP}}=512$ and $G=32$ in both 5\,GHz and 10\,GHz. On the other hand, it can also be observed that the estimation NMSE of the SOMP algorithm degrades and the gap between SOMP and GMMV-AMP becomes larger with the increase of $N_{\mathrm{AP}}$ and bandwidth. The aforementioned analysis demonstrates that both the increase of $N_{\mathrm{AP}}$ and system bandwidth can intensify the beam squint effect, which can be eliminated well by the proposed approach.}

{
	\begin{table}[t!]

		\caption{Computational Complexity of Different Channel Estimation Schemes.} 
		\label{complexity} 
		\vspace*{-5mm}
		\begin{center}
			\setlength{\tabcolsep}{0.7mm}		
			\begin{tabular}{cc}
				
				\Xhline{1.2pt}
				Schemes & Complexity\\			\Xhline{1.2pt}
				 &   $\mathcal{O} (GVKI+\frac{1}{4}I^{2}(I+1)^{2}$\\
				SOMP with $I$ layers&  $+\frac{1}{3}GI(I+1)(2I+1)$  \\ 
				&  $\frac{1}{2}GKI(I+1)+VKI)$  \\\hline
				MMV-AMP with $T_{0}$ iterations  &     $\mathcal{O} (GVKT_{0}+GN_{\mathrm{AP}}K)$        \\ \hline
				MMV-LAMP with $T$ layers &   $\mathcal{O} (GVKT+GN_{\mathrm{AP}}K)$      \\ \hline
				GMMV-AMP with $T_{0}$ iterations   &     $\mathcal{O} (GVKT_{0}+GN_{\mathrm{AP}}K)$     \\ \hline
				Proposed GMMV-LAMP with $T$ layers &   $\mathcal{O} ((GVK+K^{2})T+GN_{\mathrm{AP}}K)$     \\ \hline
				
				\Xhline{1.2pt}
			\end{tabular}
		\end{center}
		
	\end{table}

}

{We further investigate the computational complexity. In the case of offline training, the computational complexity is not a major concern. Therefore, we focus on the computational complexity in the testing stage. The computational complexity of all candidate channel estimation approaches is compared in Table~\ref{complexity}. The GMMV-AMP and MMV-AMP algorithms share similar complexity under identical numbers of iterations. Moreover, when $T_{0}=T$, MMV-LAMP, GMMV-AMP, and MMV-AMP have similar complexity as the operations in one layer/iteration are similar. The extra computational complexity of the proposed GMMV-LAMP compared with other candidates comes from the fully-connected layers in each layer. Since $GV\gg K$, the computational complexity of the GMMV-LAMP is dominated by $GVKT$. Based on the discussion above and the fact that $T_{0}$ is typically much larger than $T$, we can conclude that the proposed GMMV-LAMP network has much lower computational complexity than the conventional GMMV-AMP algorithm. Finally, the SOMP algorithm requires several operations such as correlation, project subspace and update residual operations. Therefore, the proposed network outperforms the SOMP algorithm since the complexity of SOMP increases with the fourth power of $I$.}

\begin{figure}[b!]
	
	\centering
	\includegraphics[scale=0.56]{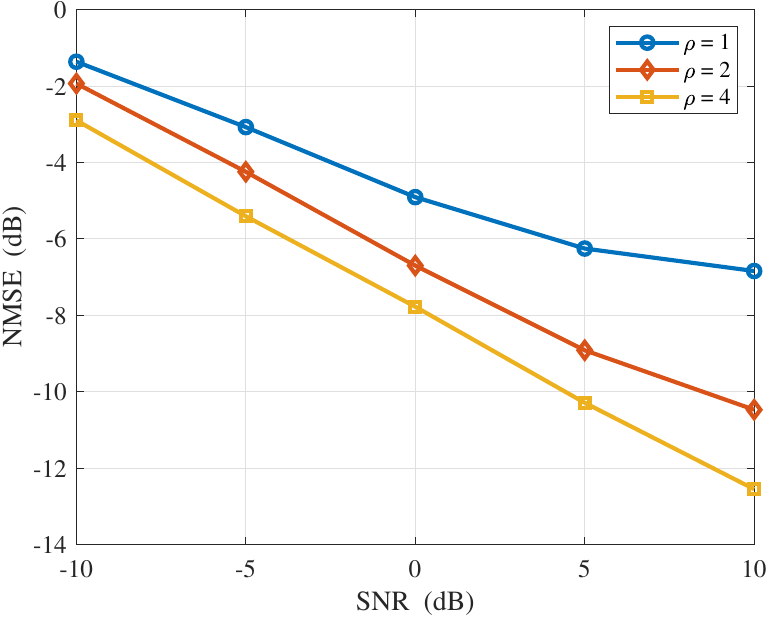}
	\vspace*{-2mm}	
	\caption{Estimation NMSE given three different dictionary redundant factors $\rho$.}
	\label{different-rho}
	\vspace*{-2mm}	
\end{figure}

\begin{figure}[h!]
	\vspace*{-5mm}	
	\centering
	\includegraphics[scale=0.36]{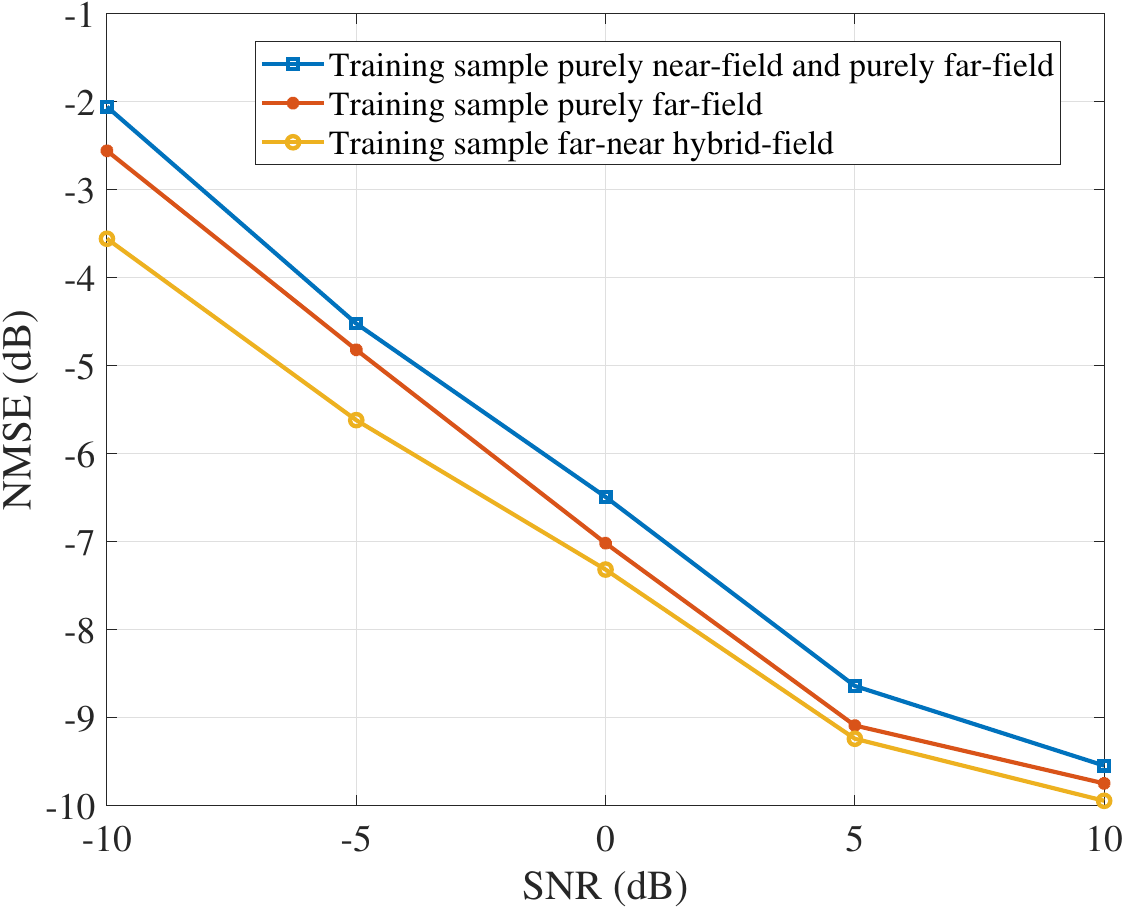}
	\vspace*{-2mm}	
	\caption{Estimation NMSE given different training datasets.}
	\label{different-sample}
\end{figure}

The influence of the redundant factor $\rho$ of $\mathbf{D}_{\rm AD}$ on the achievable CE performance for the GMMV-LAMP is illustrated in Fig.~\ref{different-rho}. Compared with the non-redundancy DFT WRD of $\rho = 1$, utilizing the redundant DFT WRDs of $\rho = 2$ significantly improves the CE accuracy, in particular achieving more than 4\,dB reduction in the NMSE at the SNR of 10\,dB. Utilizing the redundant DFT WRDs of $\rho = 4$ further brings about a 2\,dB reduction in the NMSE over the case of $\rho = 2$. {However, a larger $\rho$ also indicates that there are more trainable parameters, e.g., the dimension of $\mathbf{B}_{t}[k]$ is proportional to $\rho$. Therefore, when the number of AP antennas $N_{\mathrm{AP}}$ is small, we can employ a larger $\rho$ such as $\rho=4$. Nevertheless, if the number of antennas equipped at the AP further increases, the spatial resolution of the array has also been improved, so we can utilize a smaller $\rho$ to obtain satisfactory estimation accuracy.}

\subsubsection{Hybrid Near- and Far- Field Scenario}

Fig.~\ref{different-sample} illustrates the influence of different datasets on the achievable CE accuracy of the proposed GMMV-LAMP network in the hybrid near- and far- field scenario, where $V\! =\! 1280$ data-driven WRD is considered. The legend `Training sample purely far-field' means the purely far-field dataset. `Training sample far-near hybrid-field’ means that the CSI samples in the utilized dataset are composed of 3 far-field and 3 near-field scatterers, and `Training sample purely near-field and purely far-field' means that a mixed dataset is utilized, where the CSI samples are randomly selected from the purely far-field dataset and purely near-field dataset. As expected, the GMMV-LAMP achieves its best performance when the hybrid near- and far- field dataset is employed, since the real system is a hybrid near- and far- field one. The performance losses for the other two cases are clearly attributed to the model mismatch of the purely far-field and purely near-field CSI.

\begin{figure}[b!]
\begin{center}
\includegraphics[width=0.48\textwidth]{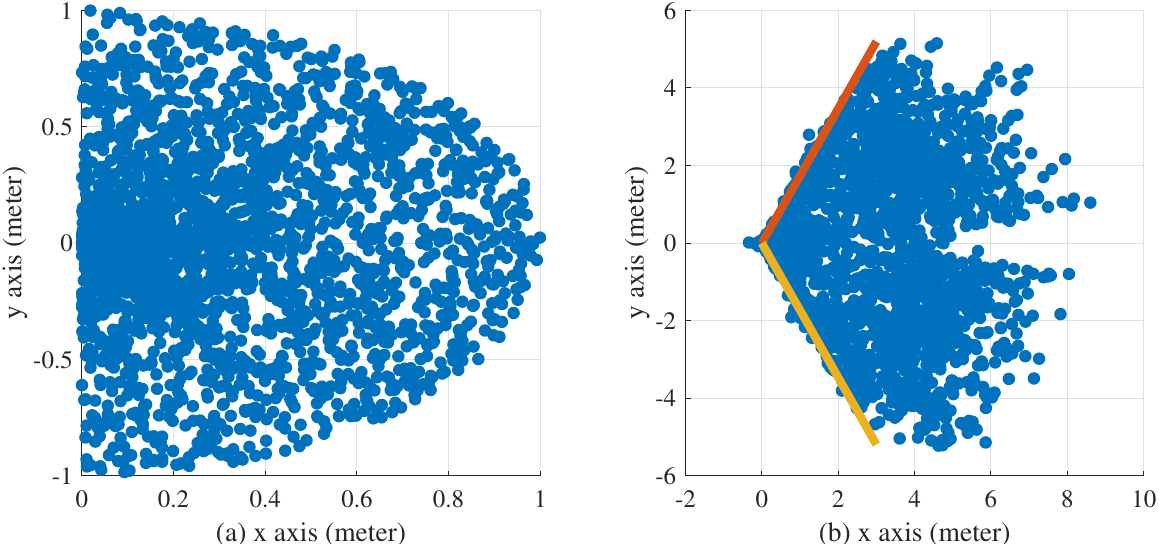}
\end{center}

\caption{Locations of the columns in the data-driven WRD: (a)~the untrained WRD, and (b)~the WRD after training.}
\label{Codebook-trained} 

\end{figure}

\begin{figure}[h!]
\vspace*{-4mm}
		\centering
		\includegraphics[scale=0.36]{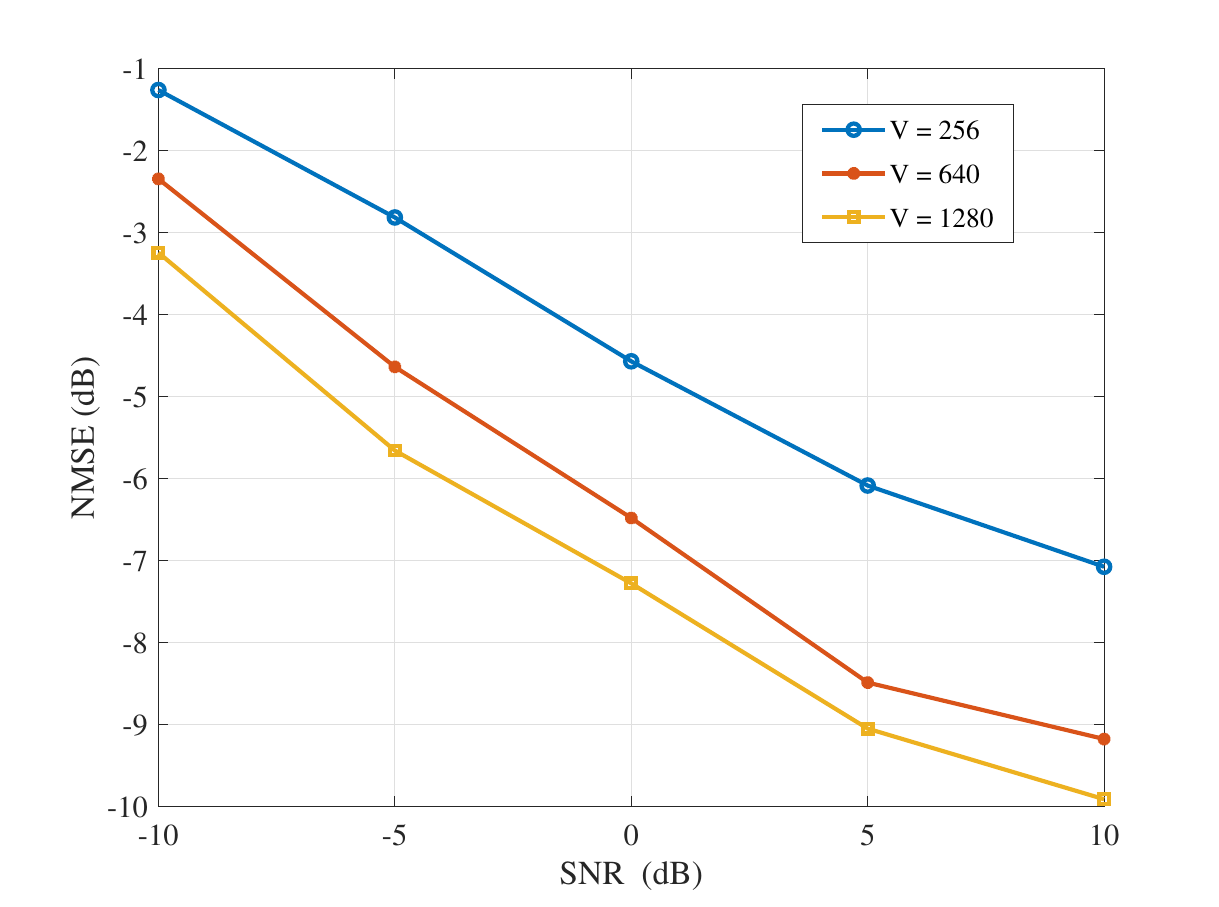}

		\caption{Estimation NMSE with different numbers of the columns $V$ in the data-driven WRD.}
		\label{different-codebook}
	\vspace*{-4mm}	

\end{figure}

To evaluate the effectiveness of the data-driven WRD, Fig.~\ref{Codebook-trained} illustrates the locations of columns in the data-driven WRD when $V=1280$ and the hybrid near- and far- field dataset are utilized. Recall that the columns in a data-driven WRD represent the steering vectors determined by the distance/AoD pairs or Cartesian coordinates. Therefore, it is natural that a WRD performs better when its coordinates of the columns are closer to the scatterers' locations, since the sparsity of support is enhanced in this case. The left subfigure depicts the Cartesian coordinates of the columns in the untrained data-driven WRD, while the right subfigure shows the Cartesian coordinates' distribution of the data-driven WRD after training. It can be observed that the trained WRD converges to a conical area within the AoDs of $\left[-\frac{\pi}{3},\, \frac{\pi}{3}\right]$, where the red and yellow lines correspond to the AoDs of $-\frac{\pi}{3}$ and $\frac{\pi}{3}$, respectively. This demonstrates that the data-driven WRDs can learn a better sparse representation from the training dataset adaptively.

The impact of the number of columns $V$ in the data-driven WRD on the achievable CE performance is investigated in Fig.~\ref{different-codebook}. It can be seen that the CE accuracy improves as $V$ increases, because a larger WRD can learn the sparse representation of CSI with more degrees of freedom. However, the performance gain by increasing $V$ from 640 to 1280 is not as pronounced as that obtained by increasing $V$ from 256 to 640. This can be attributed to the leakage issue of the off-grid scatterers, similar to increasing the redundant factor $\rho$ in Fig.~\ref{different-rho}. { In this case, the value of $V$ for data-driven WRDs should be carefully selected according to the tradeoff between complexity and performance. Based on the analysis above, taking the value of $V$ between $4N_{\mathrm{AP}}$ and $10N_{\mathrm{AP}}$ can guarantee a good estimation performance without bringing excessive computational complexity.}
\begin{figure}[b!]
	
	\vspace*{-4mm}
	\centering
	\includegraphics[scale=0.45]{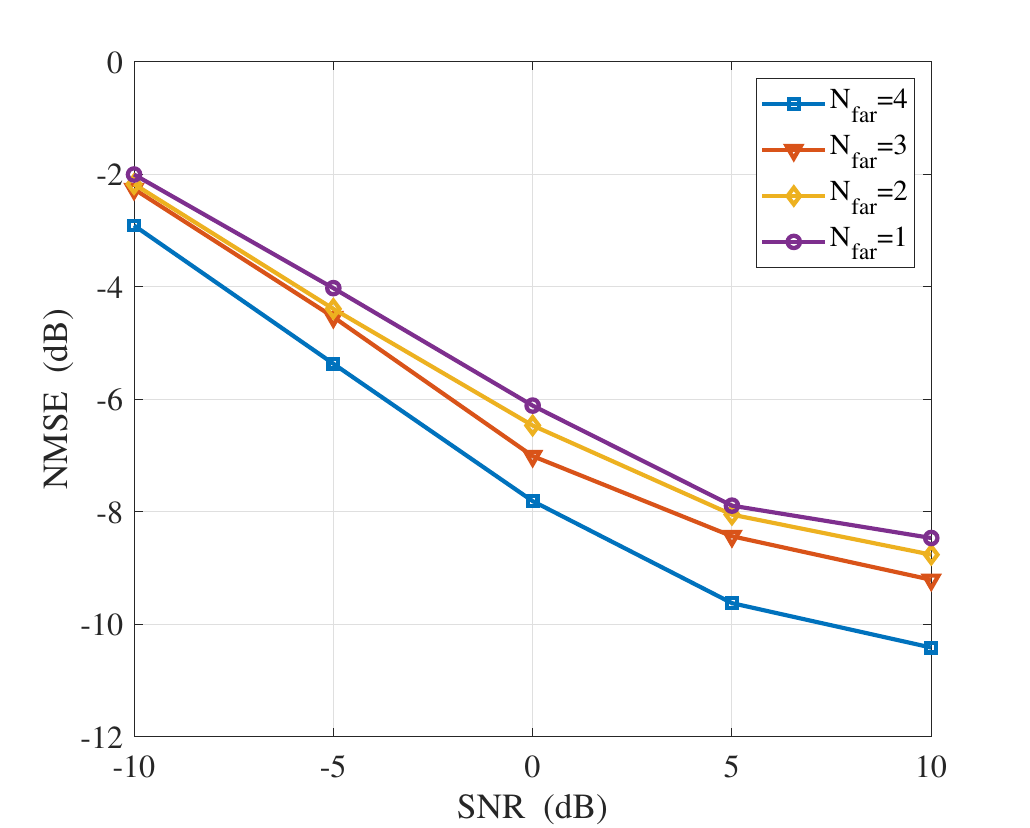}
	\vspace*{-2mm}
	\caption{Estimation NMSE with different numbers of far-field scatterers $L_f$ in the hybrid near- and far- field scenario.}
	\label{different-far}
	\vspace*{-4mm}
	
\end{figure}

\begin{figure}[h!]

	\centering
	\includegraphics[scale=0.3]{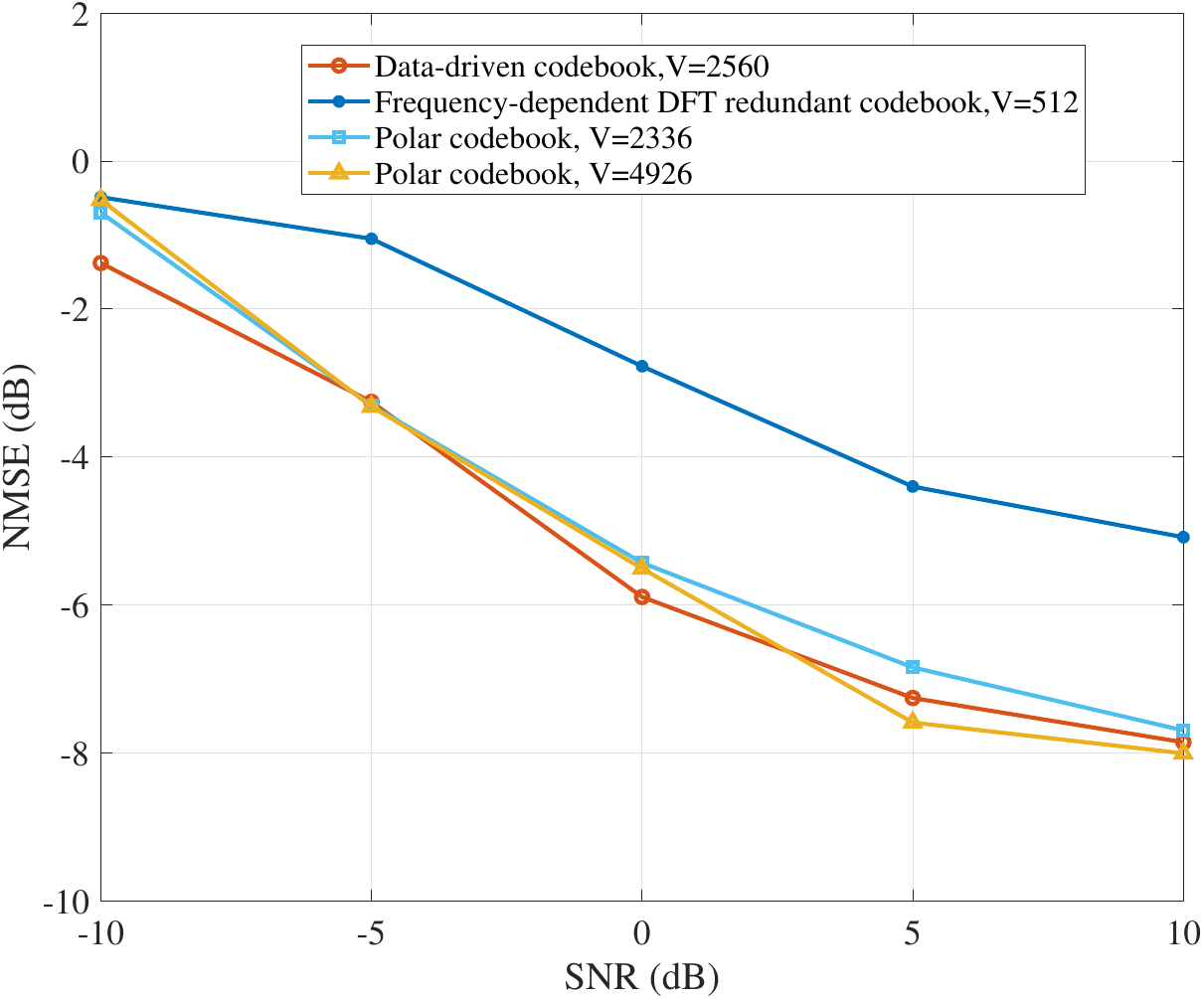}
	
	\caption{Estimation NMSE with different WRDs in the purely near-field scenario.}
	\label{NMSE_near}
	\vspace*{-2mm}	
\end{figure}

To evaluate the robustness of the proposed data-driven WRD in the hybrid near- and far- field scenario, Fig.~\ref{different-far} shows the CE NMSE when the proportion of far-field scatterers and near-field scatterers varies while fixing the number of all the scatterers to $L = 5$ and using the $V\! =\! 768$ data-driven WRD. As the number of far-field scatterers $L_f$ increases, the CE accuracy improves. Specifically, for the case of $L_f=4$ far-field scatterers and $L_n=1$ near-field scatterer, the CE accuracy is the best and the NMSE is smaller than -10\,dB at the SNR of 10\,dB. By contrast, when only 1 scatterer is located in the far-field region but there are 4 near-field scatterers in the near-field area, the NMSE is around 2\,dB worse than the best case, but the CE accuracy is still acceptable in this worst case. The aforementioned results demonstrate that the proposed approach can handle different hybrid near- and far- field scenarios.

\begin{figure}[b!]
\vspace*{-5mm}
	\centering
	\includegraphics[scale=0.5]{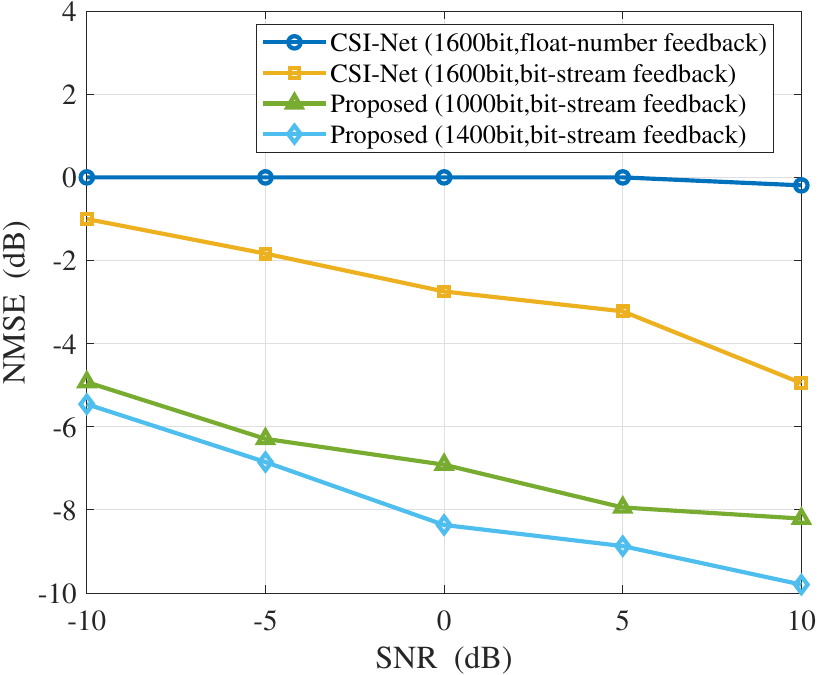}
\vspace*{-3mm}
	\caption{Reconstruction NMSE for CSI-Net \cite{8322184} and our CSI-ResNet in the hybrid near- and far- field scenario.}
	\label{feedback_hybrid}
\vspace*{-4mm}
\end{figure}

\subsubsection{Purely Near-Field Scenario}

The CE performance obtained using different WRDs in a purely near-field scenario is shown in Fig.~\ref{NMSE_near}. The DFT WRD attains the worst CE performance because there exists severe sparsity loss when the near-field scatterers are transformed into the virtual angular domain. The simulation results indicate that the proposed data-driven WRD can achieve comparable performance with the polar-domain WRD \cite{cui2022channel}, which demonstrates the effectiveness of our data-driven WRD. It is also worth noting that even when $V$ of the polar-domain WRD is doubled, there is no noticeable performance enhancement.

\subsection{Bit-vector CSI Feedback}\label{S4.3}

The results of Fig.~\ref{feedback_hybrid} verify the effectiveness of the proposed CSI-ResNet in the hybrid near- and far- field scenario where the $V\! =\! 1280$ data-driven WRD is used in CE. The legend `float-number feedback' means that the output is directly compressed as a floating-point-number vector, and `bit-vector feedback' means that the output is first compressed as a floating-point-number vector with a larger dimension and then quantized as a bit vector. It can be seen that with $N_f=1600$ feedback bits, when the benchmark CSI-Net \cite{8322184} outputs floating-point-number vectors, the reconstruction NMSE is much worse than that of the bit-vector feedback policy, which demonstrates the effectiveness of the bit-vector CSI feedback approach. {Moreover, our proposed CSI-ResNet with $N_f=1000$ feedback bits achieves significantly better reconstruction accuracy than CSI-Net with $N_f=1600$ feedback bits. Around 1\,dB further reduction in the NMSE can be attained by our CSI-ResNet with $N_f=1400$ feedback bits.}

\begin{figure}[t!]
	\begin{center}
		\includegraphics[width=0.48\textwidth]{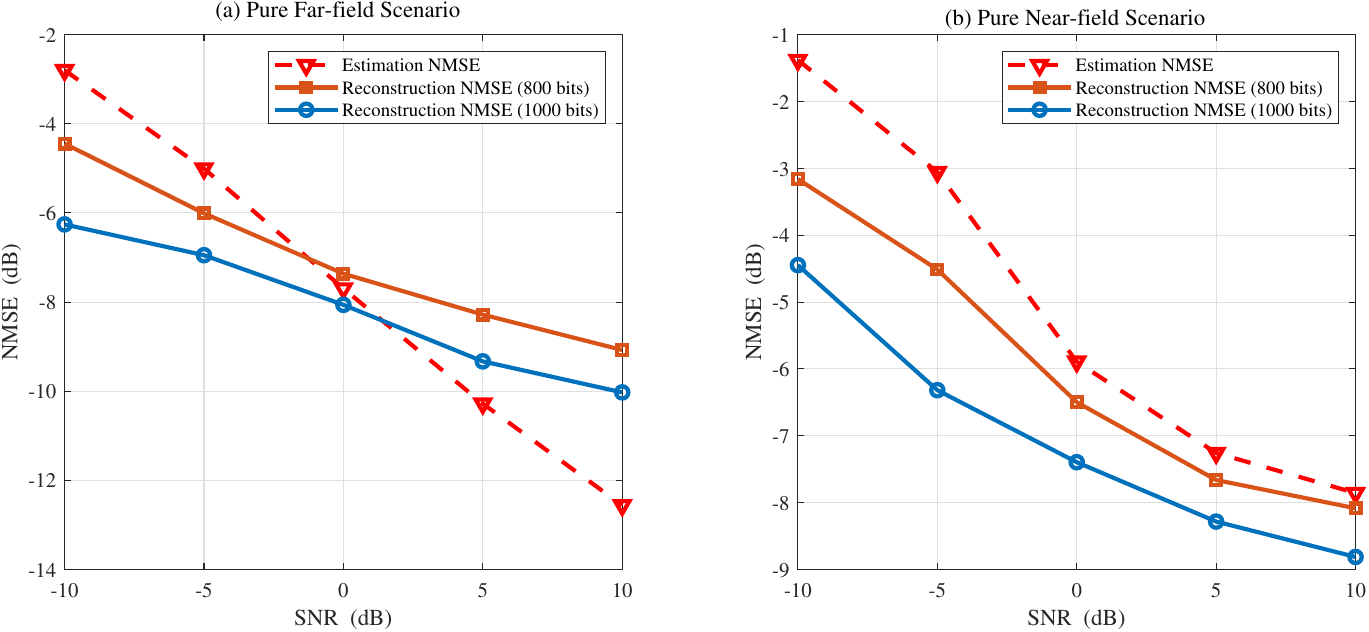}
	\end{center}
		\vspace*{-4mm}
	\caption{Reconstruction NMSE for our CSI-ResNet in purely far-field scenario and purely near-field scenario.}
	\label{feedback_all} 
	\vspace*{-4mm}
\end{figure}

Figure.~\ref{feedback_all} evaluates the CSI feedback performance of CSI-ResNet in the purely far-field and purely near-field scenarios, respectively. It should be noted that the $\rho=4$ DFT WRD and $V=2560$ data-driven WRD are used in CE for the aforementioned scenarios, respectively. It can be observed that at the high SNR range of 3 to 10\,dB, there exists the reconstruction NMSE loss in the far-field region due to the information loss during compression. However, at low SNR regions, the reconstruction NMSE presents better performance than the estimation NMSE because of the denoising effect of CSI-ResNet. In the purely near-field scenario, the denoising effect of CSI-ResNet is even more pronounced, and the reconstruction NMSE is lower than the estimation NMSE at all SNRs. In terms of the impact of feedback bits, when $N_f = 1000$ is used, the performance degrades compared with $N_f = 1400$, due to discarding more information.
\vspace*{-5mm}
\section{Conclusions}\label{S5}

In this paper, we have proposed a novel knowledge and data dual-driven CE and CSI feedback approach for downlink mmWave wideband UM-MIMO systems considering hardware imperfections. To mitigate the distortions caused by imperfect hardware, a data-driven ResNet-DQ has been proposed which is inspired by the time-domain correlation of multiple sets of oversampling samples. After de-quantization, we have proposed a DFT WRD and data-driven WRD. The former can eliminate the far-field beam squint effect and the latter can deal with both the near-field propagation effect and beam squint effects. In this way, we can obtain the exact common support across all the subcarriers in different propagation scenarios. Then, we have proposed a knowledge-driven GMMV-LAMP network based on a carefully designed shrinkage function to leverage the common support feature. Finally, we have designed a data-driven CSI feedback module called CSI-ResNet to achieve low-cost bit-vector feedback. Simulation results have indicated that the proposed approach achieves effective, accurate, and low-cost downlink CE and CSI feedback in mmWave wideband UM-MIMO systems, under a wide variety of propagation scenarios.

{\appendix[Derivation of the MMSE Denoiser]\label{AppendixA}

Consider the denoising problem for the signal model given by
\begin{equation}\label{appendix1} 
	\tilde{\mathbf{x}}=\bar{\mathbf{x}} +\mathbf{\Sigma}^{\frac{1}{2}}\mathbf{n},
\end{equation}
where $\tilde{\mathbf{x}}$ and $\bar{\mathbf{x}}\in \mathbb{C} ^{K\times 1}$ denote the noisy and noiseless signals, respectively, $\mathbf{n}\in \mathbb{C}^{K\times 1}\sim\mathcal{CN}(\mathbf{0},\mathbf{I})$ is an AWGN vector, and $\mathbf{\Sigma}=\mathrm{diag}\left(\left[\sigma^{2}[1],\sigma^{2}[2],\ldots,\sigma^{2}[K]\right]\right)$ is the diagonal matrix depicting the noise power with $\sigma^{2}[k]\ge0,\forall k\in\{1,2,\ldots,K\}$. Because the support across different subcarriers appears or disappears at the same time, $\bar{\mathbf{x}}$ follows a Bernoulli-Gaussian distribution as $(1-\gamma )\delta_{0}+\gamma p_{\mathbf{h}_{\epsilon}}$. Here, $\delta_{0}$ denotes the point mass measure at zero and $p_{\mathbf{h}_{\epsilon}}$ denotes the distribution of $\mathbf{h}_{\epsilon}\sim\mathcal{CN}(\mathbf{0},\epsilon\mathbf{I})$. 

In this way, the probability when $\tilde{\mathbf{x}}=\mathbf{x}'=\mathbf{h}_{\epsilon}+\mathbf{\Sigma}^{\frac{1}{2}}\mathbf{n}$ is $\gamma$, and the probability is $1-\gamma$ when $\tilde{\mathbf{x}}=\mathbf{\Sigma}^{\frac{1}{2}}\mathbf{n}$. According to standard estimation theory, defining $\boldsymbol{\Theta} = \mathrm{diag}\bigg(\frac{1}{\epsilon+\sigma^{2}[1]},\ldots,\frac{1}{\epsilon+\sigma^{2}[K]}\bigg)$ the mean and covariance matrix of $\mathbf{h}_{\epsilon}$ can be computed respectively as 
\begin{align}\label{appendix2} 
	\mathbb{E}[\mathbf{h}_{\epsilon}|\mathbf{x}'=\mathbf{x}] =& \epsilon\boldsymbol{\Theta}\mathbf{x},\\
	\mathbb{E}[\mathbf{h}_{\epsilon}\mathbf{h}_{\epsilon}^{\rm H}|\mathbf{x}'=\mathbf{x}] =& \epsilon \mathbf{I} - \epsilon^{2}\boldsymbol{\Theta}+\epsilon^{2}\boldsymbol{\Theta}\mathbf{x}\mathbf{x}^{\rm H}\boldsymbol{\Theta}.
\end{align}
Furthermore, we can compute the mean of $\bar{\mathbf{x}}$ as
\begin{align}\label{appendix4} 
	\mathbb{E}[\bar{\mathbf{x}}|\hat{\mathbf{x}}=\hat{\mathbf{x}}'] &= \int \bar{\mathbf{x}}p_{\mathbf{x}|\hat{\mathbf{x}}}(\bar{\mathbf{x}}=\mathbf{x}|\hat{\mathbf{x}}=\hat{\mathbf{x}}')\mathrm{d}\mathbf{x} \nonumber  \\ &\hspace*{-20mm}= \frac{1}{p_{\hat{\mathbf{x}}}}\int p_{\mathbf{x}|\hat{\mathbf{x}}}(\bar{\mathbf{x}}=\mathbf{x}|\hat{\mathbf{x}}=\hat{\mathbf{x}}')(\gamma p_{\mathbf{h}_{\epsilon}}(\mathbf{h}_{\epsilon}=\mathbf{x})+(1-\gamma)\delta_{0}(\mathbf{x}))\mathrm{d}\mathbf{x} \nonumber \\
	& = \frac{\gamma p_{\mathbf{x}'}(\mathbf{x}'=\hat{\mathbf{x}}')}{p_{\hat{\mathbf{x}}}(\hat{\mathbf{x}}=\hat{\mathbf{x}}')p_{\mathbf{x}'}(\mathbf{x}'=\hat{\mathbf{x}}')}\mathbb{E}[\mathbf{h}_{\epsilon}|\mathbf{x}'=\hat{\mathbf{x}}'],
\end{align}

By defining 
$ \phi(\hat{\mathbf{x}}) = \frac{1}{1+\frac{1-\gamma}{\gamma}e^{-\hat{\mathbf{x}}^{\rm H}\mathbf{P}\hat{\mathbf{x}}}\prod_{k=1}^{K}(1+\frac{\epsilon}{\sigma^{2}[k]}) }
$ and $
\mathbf{P} = \mathrm{diag}\left(\frac{\epsilon}{\sigma^{2}[1](\sigma^{2}[1]+\epsilon)},\ldots,\frac{\epsilon}{\sigma^{2}[K](\sigma^{2}[K]+\epsilon)}\right) ,
$
the shrinkage function $\bm{\eta}_{\rm CS}(\hat{\mathbf{x}}';\gamma,\epsilon,\mathbf{\Sigma})$ can be rewritten as 
\begin{equation}\label{appendix7} 
	\bm{\eta}_{\rm CS}(\hat{\mathbf{x}}';\gamma,\epsilon,\mathbf{\Sigma})=\mathbb{E}[\mathbf{x}|\hat{\mathbf{x}} = \hat{\mathbf{x}}']=\phi(\hat{\mathbf{x}}')\boldsymbol{\Theta}\hat{\mathbf{x}}'.
\end{equation}

It should be noted that when taking the derivative of \eqref{appendix7}, we can approximate $\phi(\hat{\mathbf{x}})$ as a constant since the dimension of $\hat{\mathbf{x}}$ is quite large, and the derivative becomes $\frac{\epsilon\phi(\hat{\mathbf{x}})}{\epsilon+\sigma^{2}[k]}$.}
\vspace*{-2mm}

\bibliographystyle{IEEEtran}


\begin{thebibliography}{10}
\providecommand{\url}[1]{#1}
\csname url@samestyle\endcsname
\providecommand{\newblock}{\relax}
\providecommand{\bibinfo}[2]{#2}
\providecommand{\BIBentrySTDinterwordspacing}{\spaceskip=0pt\relax}
\providecommand{\BIBentryALTinterwordstretchfactor}{4}
\providecommand{\BIBentryALTinterwordspacing}{\spaceskip=\fontdimen2\font plus
\BIBentryALTinterwordstretchfactor\fontdimen3\font minus
  \fontdimen4\font\relax}
\providecommand{\BIBforeignlanguage}[2]{{%
\expandafter\ifx\csname l@#1\endcsname\relax
\typeout{** WARNING: IEEEtran.bst: No hyphenation pattern has been}%
\typeout{** loaded for the language `#1'. Using the pattern for}%
\typeout{** the default language instead.}%
\else
\language=\csname l@#1\endcsname
\fi
#2}}
\providecommand{\BIBdecl}{\relax}
\BIBdecl


\bibitem{conf}
K.~Wang, Z.~Gao, Y.~Zhang, T.~Qin, R.~Na, M.~Wu, Z.~Li, ``Knowledge-driven deep learning based channel estimation for Terahertz extra-large MIMO systems,'' in \emph{Proc. IEEE 23rd Int. Conf. Commun. Technol.}, 2023.
\color{black}
\bibitem{9711564} 
M.~Vaezi, \emph{et al.}, ``Cellular, wide-area, and non-terrestrial IoT: A survey on 5G advances and the road toward 6G,'' \emph{IEEE Commun. Surveys Tuts.}, vol.~24, no.~2, pp.~1117--1174, Secondquarter, 2022.

\bibitem{8241348} 
S.~A.~Busari, \emph{et al.}, ``Millimeter-wave massive MIMO communication for future wireless systems: A survey,'' \emph{IEEE Commun. Surveys Tuts.}, vol.~20, no.~2, pp.~836--869, Secondquarter, 2018.

\bibitem{9798771} 
P.~Yang, \emph{et al.}, ``Feeling of presence maximization: mmwave-enabled virtual reality meets deep reinforcement learning,'' \emph{IEEE Trans. Wireless Commun.}, vol.~21, no.~11, pp.~10005--10019, Nov. 2022.

\bibitem{9237460} 
H.~Yang, \emph{et al.}, ``Artificial-intelligence-enabled intelligent 6G networks,'' \emph{IEEE Network}, vol.~34, no.~6, pp.~272--280, Nov./Dec. 2020.

\bibitem{BJORNSON20193} 
E.~Bj{\"o}rnson, \emph{et al.}, ``Massive MIMO is a reality--what is next?: Five promising research directions for antenna arrays,'' \emph{Digit. Signal Process.}, vol.~94, pp.~3--20, Nov. 2019.

\bibitem{add1}
L.~Xiao, S.~Li, Y.~Liu, G.~Liu, P.~Xiao and T.~Jiang, ``Error probability analysis for ultra-massive MIMO system and near-optimal signal detection,''\emph{China Commun.}, vol.~20, no.~5, pp.~1--19, May 2023.
\color{black}
\bibitem{9799524} 
X.~Cheng, Z.~Huang, and L.~Bai, ``Channel nonstationarity and consistency for beyond 5G and 6G: A survey,'' \emph{IEEE Commun. Surveys Tuts.}, vol.~24, no.~3, pp.~1634--1669, Thirdquarter 2022.

\bibitem{8844787} 
Y.~Chen, D.~Chen, T.~Jiang, and L.~Hanzo, ``Channel-covariance and angle-of-departure aided hybrid precoding for wideband multiuser millimeter wave MIMO systems,'' \emph{IEEE Trans. Commun.}, vol.~67, no.~12, pp.~8315--8328, Dec. 2019.

\bibitem{9262080} 
Y.~Chen, \emph{et al.}, ``Hybrid precoding for wideband millimeter wave MIMO systems in the face of beam squint,'' \emph{IEEE Trans. Wireless Commun.}, vol.~20, no.~3, pp.~1847--1860, Mar. 2021.

\bibitem{8354789} 
B.~Wang, \emph{et al.}, ``Spatial- and frequency-wideband effects in millimeter-wave massive MIMO systems,'' \emph{IEEE Trans. Signal Process.}, vol.~66, no.~13, pp.~3393--3406, Jul. 2018.

\bibitem{7174558} 
Z.~Gao, L.~Dai, Z.~Wang, and S.~Chen, ``Spatially common sparsity based adaptive channel estimation and feedback for FDD massive MIMO,'' \emph{IEEE Trans. Signal Process.}, vol.~63, no.~23, pp.~6169--6183, Dec. 2015.

\bibitem{7355354} 
Z.~Gao, \emph{et al.}, ``Structured compressive sensing-based spatio-temporal joint channel estimation for FDD massive MIMO,'' \emph{IEEE Trans. Commun.}, vol.~64, no.~2, pp.~601--617, Feb. 2016.

\bibitem{gao2014structured} 
Z.~Gao, L.~Dai, and Z.~Wang, ``Structured compressive sensing based superimposed pilot design in downlink large-scale MIMO systems,'' \emph{Electron Lett.}, vol.~50, no.~12, pp.~896--898, Jun. 2014.

\bibitem{9685542} 
M.~Cui and L.~Dai, ``Near-field channel estimation for extremely large-scale MIMO with hybrid precoding,'' in \emph{Proc. GLOBECOM 2021} (Madrid, Spain), Dec.~7-11, 2021, pp.~1--6.

\bibitem{9598863} 
X.~Wei and L.~Dai, ``Channel estimation for extremely large-scale massive MIMO: Far-field, near-field, or hybrid-field?'' \emph{IEEE Commun. Lett.}, vol.~26, no.~1, pp.~177--181, Jan. 2022.

\bibitem{8961111} 
M.~Ke, \emph{et al.}, ``Compressive sensing-based adaptive active user detection and channel estimation: Massive access meets massive MIMO,'' \emph{IEEE Trans. Signal Process.}, vol.~68, pp.~764--779, Jan. 2020.

\bibitem{RaoFDDCE} 
X.~Rao and V.~K.~N.~Lau, ``Distributed compressive CSIT estimation and feedback for FDD multi-user massive MIMO systems,'' \emph{IEEE Trans. Signal Process.}, vol.~62, no.~12, pp.~3261--3271, Jun. 2014.

\bibitem{7867037} 
D.~Mi, \emph{et al.}, ``Massive MIMO performance with imperfect channel reciprocity and channel estimation error,'' \emph{IEEE Trans. Commun.}, vol.~65, no.~9, pp.~3734--3749, Sep. 2017.

\bibitem{9452036} 
X.~Ma, Z.~Gao, F.~Gao, and M.~Di~Renzo, ``Model-driven deep learning based channel estimation and feedback for millimeter-wave massive hybrid MIMO systems,'' \emph{IEEE J. Sel. Areas Commun.}, vol.~39, no.~8, pp.~2388--2406, Aug. 2021.

\bibitem{8171203} 
J.~Mo, P.~Schniter, and R.~W. Heath, ``Channel estimation in broadband millimeter wave MIMO systems with few-bit ADCs,'' \emph{IEEE Trans. Signal Process.}, vol.~66, no.~5, pp.~1141--1154, Mar. 2018.

{
\bibitem{9325920} 
B.~Ning, \emph{et al.},  ``Terahertz multi-user massive MIMO with intelligent reflecting surface: Beam training and hybrid beamforming,'' \emph{IEEE Trans. Veh. Technol.}, vol.~70, no.~2, pp.~1376--1393, Feb. 2021.

\bibitem{9684752} 
B.~Shi, \emph{et al.}, ``Impact of low-resolution ADC on DOA estimation performance for massive MIMO receive array,'' \emph{IEEE Systems J.}, vol.~16, no.~2, pp.~2635--2638, Jun. 2022.}

\bibitem{8846224} 
A.~Liao, \emph{et al.}, ``Closed-loop sparse channel estimation for wideband millimeter-wave full-dimensional MIMO systems,'' \emph{IEEE Trans. Commun.}, vol.~67, no.~12, pp.~8329--8345, Dec. 2019.

\bibitem{6189758} 
P.~J.~A.~Harpe, B.~Busze, K.~Philips, and H.~de~Groot, ``A 0.47-1.6 mW 5-bit 0.5-1 GS/s time-interleaved SAR ADC for low-power UWB radios,'' \emph{IEEE J. Solid-State Circuits}, vol.~47, no.~7, pp.~1594--1602, Jul. 2012.
\bibitem{VP_book}
Y.  Eldar, A. Goldsmith, D. G\"{u}nd\"{u}z and H. V. Poor, Eds. ,{\it Machine Learning and Wireless Communications}.  (Cambridge University Press, Cambridge, UK, 2022)

\bibitem{9037126} 
X.~Ma and Z.~Gao, ``Data-driven deep learning to design pilot and channel estimator for massive MIMO,'' \emph{IEEE Trans. Veh. Technol.}, vol.~69, no.~5, pp.~5677--5682, May 2020.

\bibitem{9296779} 
H.~Hirose, T.~Ohtsuki, and G.~Gui, ``Deep learning-based channel estimation for massive MIMO systems with pilot contamination,'' \emph{IEEE Open J. Veh. Technol.}, vol.~2, pp.~67--77, Jan. 2021.

\bibitem{9246559} 
Y.~Dong, H.~Wang, and Y.-D.~Yao, ``Channel estimation for one-bit multiuser massive MIMO using conditional GAN,'' \emph{IEEE Commun. Lett.}, vol.~25, no.~3, pp.~854--858, Mar. 2021.

\bibitem{9173575} 
L.~Wan, \emph{et al.}, ``Deep learning based autonomous vehicle super resolution DOA estimation for safety driving,'' \emph{IEEE Trans. Intell. Transp. Syst.}, vol.~22, no.~7, pp.~4301--4315, Jul. 2021.

\bibitem{8322184} 
C.-K.~Wen, W.-T.~Shih, and S.~Jin, ``Deep learning for massive MIMO CSI feedback,'' \emph{IEEE Wireless Commun. Lett.}, vol.~7, no.~5, pp.~748--751, Oct. 2018.

\bibitem{8482358} 
T.~Wang, C.-K.~Wen, S.~Jin, and G.~Y.~Li, ``Deep learning-based CSI feedback approach for time-varying massive MIMO channels,'' \emph{IEEE Wireless Commun. Lett.}, vol.~8, no.~2, pp.~416--419, Apr. 2019.

\bibitem{9481880} 
Y.-C.~Lin, Z.~Liu, T.-S.~Lee, and Z.~Ding, ``Deep learning phase compression for MIMO CSI feedback by exploiting FDD channel reciprocity,'' \emph{IEEE Wireless Commun. Lett.}, vol.~10, no.~10, pp.~2200--2204, Oct. 2021.

\bibitem{9662381} 
M.~Chen, \emph{et al.}, ``Deep learning-based implicit CSI feedback in massive MIMO,'' \emph{IEEE Trans. Commun.}, vol.~70, no.~2, pp.~935--950, Feb. 2022.

\bibitem{9446900} 
J.~Wang, \emph{et al.}, ``Compressive sampled CSI feedback method based on deep learning for FDD massive MIMO systems,'' \emph{IEEE Trans. Commun.}, vol.~69, no.~9, pp.~5873--5885, Sep. 2021.

\bibitem{9814463} 
M.~Wu, \emph{et al.}, ``Deep learning-based hybrid precoding for FDD massive MIMO-OFDM systems with a limited pilot and feedback overhead,'' in \emph{Proc. ICC 2022 Workshops} (Seoul, South Korea), May~16-20, 2022, pp.~318--323.

\bibitem{DLbeamTraining1} 
K.~Ma, \emph{et al.}, ``Deep learning assisted calibrated beam training for millimeter-wave communication systems,'' \emph{IEEE Trans. Commun.}, vol.~69, no.~10, pp.~6706--6721, Oct. 2021.

\bibitem{DLbeamTraining2} 
Y.~Zhuo, Z.~Sha, Z.~Wang, and S.~Chen, ``Simultaneous multi-beam training for millimeter-wave communication
system,'' \emph{IEEE Trans. Veh. Technol.}, vol.~71, no.~10, pp.~10631--10645, Oct. 2022.

\bibitem{DLbeamPrediction} 
K.~Ma, \emph{et al.}, ``Deep learning assisted mmWave beam prediction for heterogeneous networks: A dual-band fusion approach,'' \emph{IEEE Trans. Commun.}, vol.~71, no.~1, pp.~115--130, Jan. 2023.

\bibitem{9252919} 
Y.~Liu, \emph{et al.}, ``Situation-aware resource allocation for multi-dimensional intelligent multiple access: A proactive deep learning framework,'' \emph{IEEE J. Sel. Areas Commun.}, vol.~39, no.~1, pp.~116--130, Jan. 2021.

\bibitem{DLdetection1} 
D.~He, \emph{et al.}, ``Deep learning-assisted TeraHertz QPSK detection relying on single-bit quantization,'' \emph{IEEE Trans. Commun.}, vol.~69, no.~12, pp.~8175--8187, Dec. 2021.

\bibitem{DLdetection2} 
Z.~Huang, \emph{et al.}, ``Autoencoder with fitting network for Terahertz wireless communications: A deep learning approach,'' \emph{China Commun.}, vol.~19, no.~3, pp.~172--180, Mar. 2022.

{
	\bibitem{hht_magazine}
	H.~He, \emph{et al.}, ``Model-driven deep learning for physical layer communications,'' \emph{IEEE Wireless Commun.}, vol.~26, no.~5, pp.~77--83, Oct. 2019.}



\bibitem{8715473} 
Y.~Wei, \emph{et al.}, ``An AMP-based network with deep residual learning for mmWave beamspace channel estimation,'' \emph{IEEE Wireless Commun. Lett.}, vol.~8, no.~4, pp.~1289--1292, Aug. 2019.

\bibitem{9847603} 
L.~V.~Nguyen, D.~H.~N.~Nguyen, and A.~L.~Swindlehurst, ``Deep learning for estimation and pilot signal design in few-bit massive MIMO systems,'' \emph{IEEE Trans. Wireless Commun.}, vol.~22, no.~1, pp.~379--392, Jan. 2023.
	
\bibitem{9685707} 
W.~Jin, \emph{et al.}, ``Adaptive channel estimation based on model-driven deep learning for wideband mmWave systems,'' in \emph{Proc. GLOBECOM 2021} (Madrid, Spain), Dec.~7-11, 2021, pp.~1--6.

\bibitem{9171358} 
Y.~Sun, \emph{et al.}, ``AnciNet: An efficient deep learning approach for feedback compression of estimated CSI in massive MIMO systems,'' \emph{IEEE Wireless Commun. Lett.}, vol.~9, no.~12, pp.~2192--2196, Dec. 2020.	

\bibitem{greyboxM1} 
X.~Hong and S.~Chen, ``A new RBF neural network with boundary value constraints,'' \emph{IEEE Trans. Systems, Man, and Cybernetics, Part B}, vol.~39, no.~1, pp.~298--303, Feb. 2009.

\bibitem{greyboxM2} 
S.~Chen, X.~Hong, and C.~J.~Harris, ``Grey-box radial basis function modelling: The art of incorporating
prior knowledge,'' in \emph{Proc. SSP Workshop 2009} (Cardiff, UK), Aug.~31-Sep.~3, 2009, pp.~377--380.

\bibitem{greyboxM3} 
S.~Chen, X.~Hong, and C.~J.~Harris, ``Grey-box radial basis function modelling,'' \emph{Neurocomputing}, vol.~74, no.~10, pp.~1564--1571, May 2011.	
	
\bibitem{DualA1} 
H.~He, C.-K.~Wen, S.~Jin, and G.~Y.~Li, ``Deep learning-based beamspace channel estimation in mmWave massive MIMO systems,'' \emph{IEEE Wireless Commun. Lett.}, vol.~7, no.~5, pp.~852--855, Oct. 2018.
	
\bibitem{DualA2} 
Q.~Ji, C.~Zhu, and X.~Guo, ``MmWave MIMO: An LAMP-based network with deep residual learning combining the prior channel information for beamspace sparse channel estimation,'' in \emph{Proc. ICCT 2021} (Tianjin, China), Oct.~13-16, 2021, pp.~279--284.


\bibitem{add_CERIS} 
S.~Zeng, \emph{et al.}, ``Hybrid driven learning for channel estimation in intelligent reflecting surface aided millimeter wave communications,'' \emph{arXiv preprint}, arXiv: 2305.19005, 2023.
\color{black}
\bibitem{950789} 
M.~Valkama, M.~Renfors, and V.~Koivunen, ``Advanced methods for I/Q imbalance compensation in communication receivers,'' \emph{IEEE Trans. Signal Process.}, vol.~49, no.~10, pp.~2335--2344, Oct. 2001.

\bibitem{svchannel} 
T.~S.~Rappaport, R.~W.~Heath, R.~C.~Daniels, and J.~N.~Murdock, \emph{Millimeter Wave Wireless Commun.}. Pearson Education, 2014.

\bibitem{cuibeamsplit}
M.~Cui, L.~Dai, R.~Schober, L.~Hanzo, ``Near-field wideband beamforming for extremely large antenna arrays,'' \emph{arXiv preprint}, arXiv: 2109.11054, 2021.
\color{black}
\bibitem{wang2019beamsquint} 
B.~Wang, \emph{et al.}, ``Beam squint and channel estimation for wideband mmWave massive MIMO-OFDM systems,'' \emph{IEEE Trans. Signal Process.}, vol.~67, no.~23, pp.~5893--5908, Dec. 2019.

\bibitem{10130575} 
Z.~Li, Z.~Gao, and T.~Li, ``Sensing user's channel and location with Terahertz extra-large reconfigurable intelligent surface under hybrid-field beam squint effect,'' \emph{IEEE J. Sel. Topics Signal Process.}, vol.~17, no.~4, pp.~893--911, Jul. 2023. 

\bibitem{AMP-L1} 
D.~L.~Donoho, A.~Maleki, and A.~Montanari, ``Message passing algorithms for compressed sensing,'' \emph{Proc. Nat. Acad. Sci.}, vol.~106, no.~45, pp.~18914--18919, Nov. 2009.

\bibitem{7934066} 
M.~Borgerding, P.~Schniter, and S.~Rangan, ``AMP-inspired deep networks for sparse linear inverse problems,'' \emph{IEEE Trans. Signal Process.}, vol.~65, no.~16, pp.~4293--4308, Aug. 2017.

\bibitem{shrinkage_function} 
L. Liu and W. Yu, ``Massive connectivity with massive MIMO—Part I: Device activity detection and channel estimation,'' \emph{IEEE Trans. Signal Process.}, vol.~66, no.~11, pp.~2933--2946, Jun. 2018.

\bibitem{Tranf} 
Y.~Wang, \emph{et al.}, ``Transformer-empowered 6G intelligent networks: From massive MIMO processing to semantic communication,'' \emph{IEEE Wireless Commun.} vol.~30, no.~6, pp.~127--135, Dec. 2023.

\bibitem{cui2022channel} 
M.~Cui and L.~Dai, ``Channel estimation for extremely large-scale MIMO: Far-field or near-field?'' \emph{IEEE Trans. Commun.}, vol.~70, no.~4, pp.~2663--2677, Apr. 2022.
  
\bibitem{UMMIMO_antenna} 
A.~Amiri, S.~Rezaie, C.~N.~Manch\'on and E.~de~Carvalho, ``Distributed receiver processing for extra-large MIMO arrays: A message passing approach,'' \emph{IEEE Trans. Wireless Commun.}, vol.~21, no.~4, pp.~2654-2667, Apr.~2022.
\color{black}


\end{thebibliography}

\vfill

\end{document}